\documentclass[fleqn,usenatbib]{mnras}

\usepackage{newtxtext,newtxmath}

\usepackage[T1]{fontenc}
\usepackage{ae,aecompl}

\usepackage{graphicx}	
\usepackage{amsmath}	
\usepackage{amssymb}	
\usepackage{enumitem}   

\def\q{\mathbf q} 
\def\kk{\mathbf k}
\def\x{\mathbf x}

\def\s{\mathbf s}
\def\v{\mathbf v}

\def\ebar{\overline{e}}

\def\dbar{\overline{\delta}}

\def\rbar{\overline{R}}
\def\sone{\s^{\scalebox{0.6}{(1)}}}
\def\Done{D^{\scalebox{0.6}{(1)}}}
\def\Dtwo{D^{\scalebox{0.6}{(2)}}}
\def\eonebar{\overline{e}^{\scalebox{0.6}{(1)}}}
\def\sonebar{\overline{\s}^{\scalebox{0.6}{(1)}}}
\def\stwobar{\overline{\s}^{\scalebox{0.6}{(2)}}}

\def\stwo{{\s}^{\scalebox{0.6}{(2)}}}
\def\nhat{\hat{n}}

\title[Fast halo catalogues with the mass-Peak Patch method]{The mass-Peak Patch algorithm for fast generation of deep all-sky dark matter halo catalogues and its N-Body validation}

\author[G. Stein et al.]{
George Stein$^{1,2}$\thanks{E-mail: gstein@cita.utoronto.ca},
Marcelo A. Alvarez$^{3}$,
J. Richard Bond$^{2}$
\\
$^{1}$Department of Astronomy \& Astrophysics,  University of Toronto, 50 St. George St., Toronto, ON, M5S 3H4, Canada\\
$^{2}$Canadian Institute for Theoretical Astrophysics,  University of Toronto, 60 St. George St., Toronto, ON, M5S 3H8, Canada\\
$^{3}$Berkeley Center for Cosmological Physics, 341 Campbell Hall, University of California, Berkeley, Berkeley, CA, 94720
}

\date{Accepted XXX. Received YYY; in original form ZZZ}

\pubyear{2018}

\begin{document}
\label{firstpage}
\pagerange{\pageref{firstpage}--\pageref{lastpage}}
\maketitle

\begin{abstract}
We present a detailed description and validation of our massively-parallel update to the mass-Peak Patch method, a fully predictive initial-space algorithm to quickly generate dark matter halo catalogues in very large cosmological volumes.  We perform an extensive systematic comparison to a suite of N-body simulations covering a broad range of redshifts and simulation resolutions, and find that, without any parameter fitting, our method is able to generally reproduce N-body results while typically using over 3 orders of magnitude less CPU time, and a fraction of the memory cost. Instead of calculating the full non-linear gravitational collapse determined by an N-body simulation, the mass-Peak Patch method finds an overcomplete set of just-collapsed structures around a hierarchy of density-peak points by coarse-grained (homogeneous) ellipsoidal dynamics. A complete set of \textit{mass-peaks}, or halos, is then determined by exclusion of overlapping patches, and second order Lagrangian displacements are used to move the halos to their final positions and to give their flow velocities.  Our results show that the mass-Peak Patch method is well-suited for creating large ensembles of halo catalogues to mock cosmological surveys, and to aid in complex statistical interpretations of cosmological models.

\end{abstract}


\section{Introduction}
\label{sec:intro}

Future large scale structure and cosmic microwave background surveys require extensive ensembles of large-volume mock halo catalogs in order to model observables, estimate their covariance matrices, study their systematic errors, and test data analysis pipelines. Due to the size of these surveys, and the need to simulate many different cosmological parameter values and ``beyond the standard model cosmologies'', this task becomes computationally restrictive when using traditional N-body methods to generate halo catalogs. Therefore, accelerated methods for generating simulations of the large scale structure of the universe are widely desired in cosmology, and this has resulted in an introduction of many ``approximate'' methods over the last few decades. Each of these methods is built upon different approximations to the fully non-linear gravitational collapse calculated by an N-body simulation, and many use fit parameters calibrated to the N-body results. Our focus here is on the mass-Peak Patch method of \citet{1996ApJS..103....1B,1996ApJS..103...41B,1996ApJS..103...63B}, the natural outgrowth of the `peaks' picture of \citet{1986ApJ...304...15B} and the excursion set picture of \citet{1991ApJ...379..440B}, augmented by the fundamental recognition of the role of tides and strain in defining dynamical evolution in the cosmic web. Approximate methods can be divided into three main groups: \textit{stochastic}, \textit{abridged particle mesh}, and \textit{predictive}.

Stochastic methods such as EZMOCKS \citep{2015MNRAS.446.2621C}, HALOGEN \citep{2015MNRAS.450.1856A}, PATCHY \citep{2014MNRAS.439L..21K}, PThalos \citep{2002MNRAS.329..629S, 2013MNRAS.428.1036M}, QPM \citep{2014MNRAS.437.2594W}, and log-normal models \citep{1991MNRAS.248....1C,2017JCAP...10..003A}, apply a (usually stochastic) model to a density field in order to lay down halos or galaxies, commonly based on free parameters fit to N-body results. Though fast, they require calibration, and have difficulty extending to regimes (cosmology, redshifts, or higher point estimators) where they have not already been fit \citep{2018arXiv180609499C}.

The second group, which includes COLA \citep{2013JCAP...06..036T, 2015A&C....12..109H, 2016MNRAS.459.2327I} and FastPM \citep{2016MNRAS.463.2273F}, can be considered as abridged particle mesh methods. These use additional approximations on top of standard particle-mesh gravity solvers to attempt to achieve accurate gravitational evolution with far fewer timesteps than required by a pure N-body, while still using standard N-body halo finding techniques. COLA uses the N-body solver to add a residual displacement on top of solutions from Lagrangian Perturbation Theory (LPT), while FastPM uses a broadband correction applied at each time step to artificially enforce the correct linear growth on large scales. These are very accurate for approximate methods, but still require $\sim$10-40 timesteps, halo finding, and can require a significant memory overhead. 

The third group can be dubbed predictive methods, which the mass-Peak Patch method belongs to. Others in this class include PINOCCHIO \citep{2002MNRAS.331..587M, 2013MNRAS.433.2389M}, and HaloNet \citep{2018arXiv180504537B}. These methods implement physical models of gravitational collapse (or machine learned models in the case of HaloNet) to directly find halos in the initial conditions, and then use a version of Lagrangian perturbation theory to move them to their final positions\footnote{This differs from a common definition of `predictive' used in the field, where a predictive method is instead defined as any method that `finds' halos in a given density field - initial or final. Using this definition instead would then also include N-Body, COLA, FastPM, and, arguably, the second implementation of PThalos.}. PINOCCHIO, the most similar to our Peak Patch method, computes collapse times of individual dark matter particles using an ellipsoidal collapse model based on LPT, and then applies a prescription to fragment the collapsed cells into distinct objects (halos). This method still requires calibration to N-body simulations to determine free parameters in the fragmentation step. 

The \textit{mass-Peak Patch} (mPP) method described and validated here explicitly models nonlinear halo collapse and exclusion. As such, it requires essentially no additional calibration with respect to N-body simulations, as the basic ingredients of coarse-grained local ellipsoidal dynamics into the nonlinear regime, hierarchical exclusion of small-scale entities within larger-scale entities, and bulk motion and flow, are invariant. Radical variation of cosmological parameters does not alter the algorithm, nor does focusing on different halo definitions, selecting on secondary halo properties, or even choosing to concentrate on dynamical entities that are not halos. This allows modular improvements to be straightforwardly incorporated, e.g., improving flow dynamics beyond second-order LPT (2LPT) (such as \citet{2013MNRAS.435L..78K} or \citet{2016MNRAS.455L..11N}), adding non-CDM forces to ellipsoidal dynamics, or extending to all manner of ``beyond the standard model of cosmology'' situations. The mass-Peak Patch method also requires significantly less memory usage than abridged particle mesh methods, at a minimum (maximum) of 28 (84) bytes per particle, depending on if memory is the limiting step, compared to 120 bytes for FastPM and 144 bytes for COLA (when using an optimistic value of a force resolution factor, $B=2$, and a no over-allocation of memory storage ($A=1$), which is in practice needed. Using routine values of $B=3$ and $A=1.5$ increases the memory requirements to 300 and 336 bytes per particle for FastPM and COLA, respectively) \citep{2016MNRAS.463.2273F}.

Many validation exercises have shown that the mass-Peak Patch method works well, beginning with the second of the original set of mPP papers, {\it The Peak-Patch Picture of Cosmic Catalogs: II Validation} \citep{1996ApJS..103...41B}, which used the then state-of-the-art $128^3$ adaptive particle particle mesh CDM (without $\Lambda$) simulations of \citet{1991ApJ...368L..23C} and the first spherical overdensity halo finder to demonstrate with many statistical tests the accuracy of the Peak-Patch catalogs for the cluster-group system, including of the mass function and the spatial distribution. Further testing was done throughout the nineties, including on $\Lambda$CDM cosmology. The halos in the mass-Peak Patch method, oriented according to the tidal field, were the basis of the theory of the cosmic web of \citet{1996Natur.380..603B}.  

Very recently, motivated by the mock simulation requirements for the Euclid mission \citep{2011arXiv1110.3193L}, seven of the approximate methods, including mPP, were tested in detail in a set of three papers, comparing their clustering statistics and covariance matrices with those from N-body simulations constructed from the same initial conditions. \citet{2018arXiv180609477L} compared the correlation functions, \citet{2018arXiv180609497B} compared the power spectrum multipoles, and \citet{2018arXiv180609499C} compared the bispectra. These papers showed all the methods fared well, including the mPP method. For another recent performance comparison of approximate methods (not including mPP) see \citet{2015MNRAS.452..686C}, and for a detailed review of the subject see \citet{2016Galax...4...53M}. Though it may seem then that the mPP method has been validated in a modern as well as an ancient setting, these works were limited to a single redshift and simulation resolution. In this work we expand upon these validations to multiple redshifts and simulation resolutions. 

\begin{figure*}
\begin{center}
\includegraphics[width=1.0\textwidth,trim={0 0 0 0},clip]{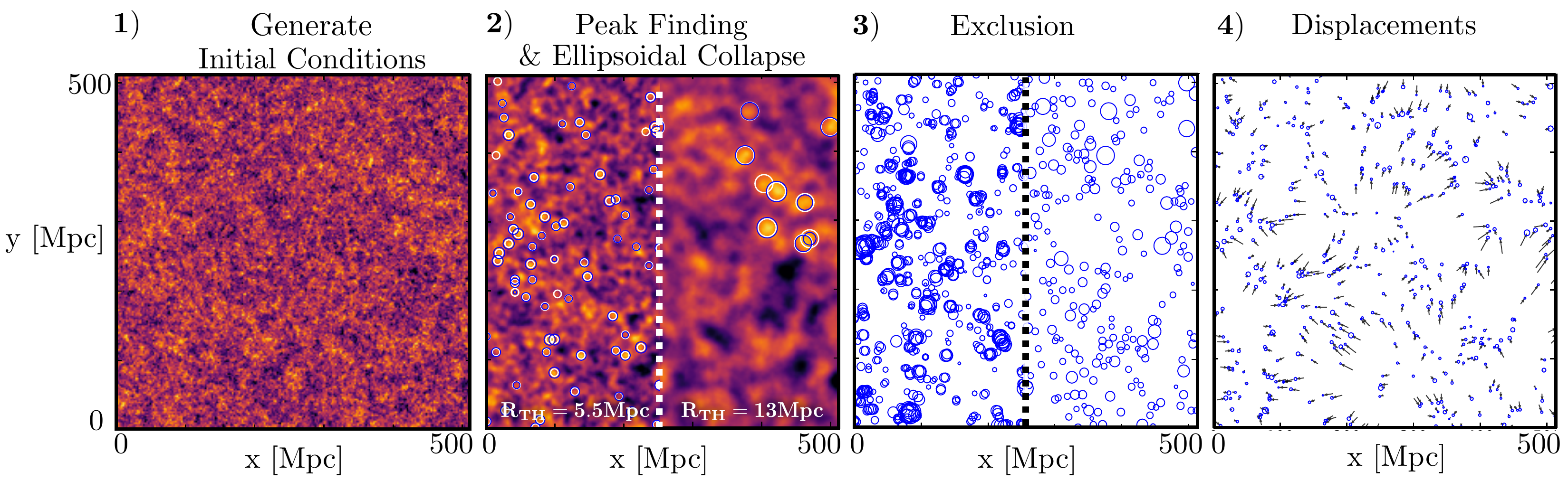}
\end{center}
\vspace{-0.3cm}
\caption{The four main steps of the mass-Peak Patch method. {\bf{1)}} Generation of a linear density field at $z=0$. {\bf{2)}} Candidate halo locations (white circles, with a radius equal to the smoothing scale they were found at) are found by top-hat smoothing the linear density field on a hierarchy of filter scales and selecting density peaks above the threshold $\delta_c \sim 1.5$. The Lagrangian radius of the collapsed region (blue circles) is then determined by adjusting the radius until the nonlinear ellipsoid trajectories, homogeneous over that radius in all three directions, have compactified to near-zero volume by the redshift in question. Halos in a slice thickness equal to the filter size are shown, and the left (right) half of the image shows the result of smoothing with a top-hat filter size equal to 5.5 Mpc (13 Mpc). Note that some density peaks above the linear density threshold do not end up collapsing, a consequence of the role of tides and strain in the gravitational collapse of a halo. {\bf{3)}} Exclusion is performed to trim overlapping patches in order to create a final complete set of disconnected ordered mass-peaks (halos). The left side of the image shows the full catalogue before exclusion, while on the right side we have performed exclusion. {\bf{4)}} Halos are moved to their final positions using second order Lagrangian perturbation theory. Black arrows show the displacements from the initial positions, and the halo size shown is $R_{200,M}$.}

\label{fig:algorithm}
\end{figure*}

In \S \ref{sec:method} we present the updated and massively-parallel version of the mass-Peak Patch method of \citet{1996ApJS..103....1B}, highlighting and testing the inherent choices needed to run the large-scale simulations required for the mocks of current and future data. In \S \ref{sec:comparisons} we validate mass-Peak Patch halo catalogues against those from a suite of N-body runs at various redshifts and simulation resolutions by comparing with a variety of summary statistics, including novel halo-proximity measures. In \S \ref{sec:discussion} we summarize our findings and discuss their implications along with the many important applications of the method already underway. Finally, in Appendix~\ref{sec:appendixa} we discuss in detail the performance and parallelization scheme of the updated Peak Patch method. 

\section{The Mass-Peak Patch Method}
\label{sec:method}

We generate halo catalogs with the mass-Peak Patch approach, originally described in \citet{1996ApJS..103....1B,1996ApJS..103...41B,1996ApJS..103...63B} (hereafter refereed to as BM), and recently used to create large synthetic mocks of the extragalactic microwave sky\footnote{mocks.cita.utoronto.ca} (Alvarez et al. 2018, in prep.), mocks of the carbon monoxide line emission from high redshift galaxies \citep{2018arXiv180807487T}, and to create covariance matrices of clustering statistics  \citep{2018arXiv180609477L, 2018arXiv180609497B, 2018arXiv180609499C}. In this new version, we have adapted the method to run on massively parallel computing architectures, and performed systematic studies of the accuracy and parameter choices required for application to large scale structure surveys. \citet{1996ApJS..103....1B} provides the physical motivation and analytical derivations, while here we provide a brief summary of our new implementation and the parameters needed to run a given simulation.

The mass-Peak Patch method can be thought of as a dark matter halo finder that operates on the initial (Lagrangian) density field, which then uses Lagrangian perturbation theory to displace the halos to their final (Eulerian) positions. The gravitational collapse of halos is approximated by solving the set of ordinary differential equations of homogeneous ellipsoid collapse, which depend only on the strain tensor of the linear displacements averaged over a corresponding spherical region in Lagrangian space. The four main steps of the method can be understood through Figure~\ref{fig:algorithm} and are described in more detail in the following list:

\begin{enumerate}[label=\textbf{\arabic*})]
\item Generation of the initial overdensity field $\delta(\q) = \frac{\rho(\q) - \overline{\rho}}{\overline{\rho}}$, where $\q$ is the initial (Lagrangian) position. This is accomplished by first generating a white noise field on a three dimensional lattice by drawing each lattice value from a Gaussian distribution with unit variance and a mean of 0, and then convolving with the linear matter power spectrum for the given cosmological parameters of the simulation. As the mass-Peak Patch method works on a discrete periodic cubic lattice, we perform many operations in the Fourier domain, where the Fourier transform of the matter overdensity field, $\delta(\kk)$, is defined through
\begin{equation}
\delta(\q)= \frac{1}{(2 \pi)^3} \int d\kk\ e^{i \kk \cdot \q} \delta(\kk) .
\end{equation}
Throughout this paper we use the same notation for variables in either domain, and simply denote operations done in real space with the Lagrangian $\q$ (or Eulerian $\x$) variable, and those done in Fourier space by the use of $\kk$. 

The first and second order linear displacements, $\sone(\q)$ and $\stwo(\q)$, are then calculated from the density field through 
\begin{equation}
\sone(\q)=\frac{\Done}{(2\pi)^3} \int d\kk\ i \frac{\kk}{k^2} e^{i \kk \cdot \q} \delta(\kk)
\end{equation}
and 
\begin{align}
\stwo(\q) &=-\frac{\Dtwo}{(2\pi)^3} \int d\kk\ i \frac{\kk}{k^2} e^{i \kk \cdot \q} F(\kk), \\
\mathrm{where}\ & F(\q)= \sum_{i>j}[\sone_{i,i}(\q)\sone_{j,j}(\q)-
\sone_{i,j}(\q)\sone_{i,j}(\q)],
\end{align}
$\sone_{i,j}(\q)=\frac{d\sone_i(\q)}{dq_j}$ (so $\sone_{i,j}(\kk)=-k_ik_j\delta(\kk)/k^2$), and $\Done$ and $\Dtwo$ are the first and second order linear growth factor, respectively. By Fourier transforming to configuration space and performing the multiplications necessary sequentially,  we are able to maintain the memory usage of the code at seven floats per mass element ($\delta(\q)$, $\sone(\q)$, $\stwo(\q)$) at the expense of only one additional Fourier transform. All fields are expressed in terms of their values at a redshift of 0. 

\item The location of potential dark matter halo candidates are found by smoothing the density field on a number of filter scales, $R_f$, to find regions above a certain density threshold by
\begin{align}
\delta(\q;R_f) &= \frac{1}{(2\pi)^3} \int d\kk\ e^{i \kk \cdot \q} \delta(\kk) W_f(\kk; R_f),
\end{align}
where $f$ denotes the type of filtering used ($f=G\to$ Gaussian, $f=TH\to$ Top Hat, etc.). The size (mass) of the halo is found by solving a set of homogeneous ellipsoidal collapse equations to determine if the given region will gravitationally collapse by the redshift in question. The collapse redshift is determined by the linear strain $\eonebar_{ij}\equiv (\partial{\sonebar_i}/\partial{q_j}+\partial{\sonebar_j}/\partial{q_i})/2$, averaged over a sphere centered on each region of interest (see Section~\ref{sec:ellipsoidalcollapse} for details).

\item Overlapping regions of Lagrangian space belonging to multiple halos are dealt with through exclusion, which can be of various physically-motivated varieties as described in Section~\ref{sec:exclusion}.

\item Halos are then displaced to their final (Eulerian) position $\x_h$, at redshift $z$, using $\x_h(z)=\q_h + \Done(z_h)\sonebar + \Dtwo(z_h)\stwobar$, where $\sonebar$ and $\stwobar$ are the first and second order displacement fields averaged over the volume of the halo. We use the standard approximation $\Dtwo(z) = -\frac{3}{7}D(z)^2\Omega_m^{-1/143}$ \citep{1995A&A...296..575B}. The velocity of each halo is determined by $\v_h=a_h[\dot{\Done}(z_h)\sonebar+\dot{\Dtwo}(z_h)\stwobar].$
\end{enumerate}

\begin{figure*}
\begin{center}
\includegraphics[width=1.\textwidth,trim={0 0 0 0},clip]{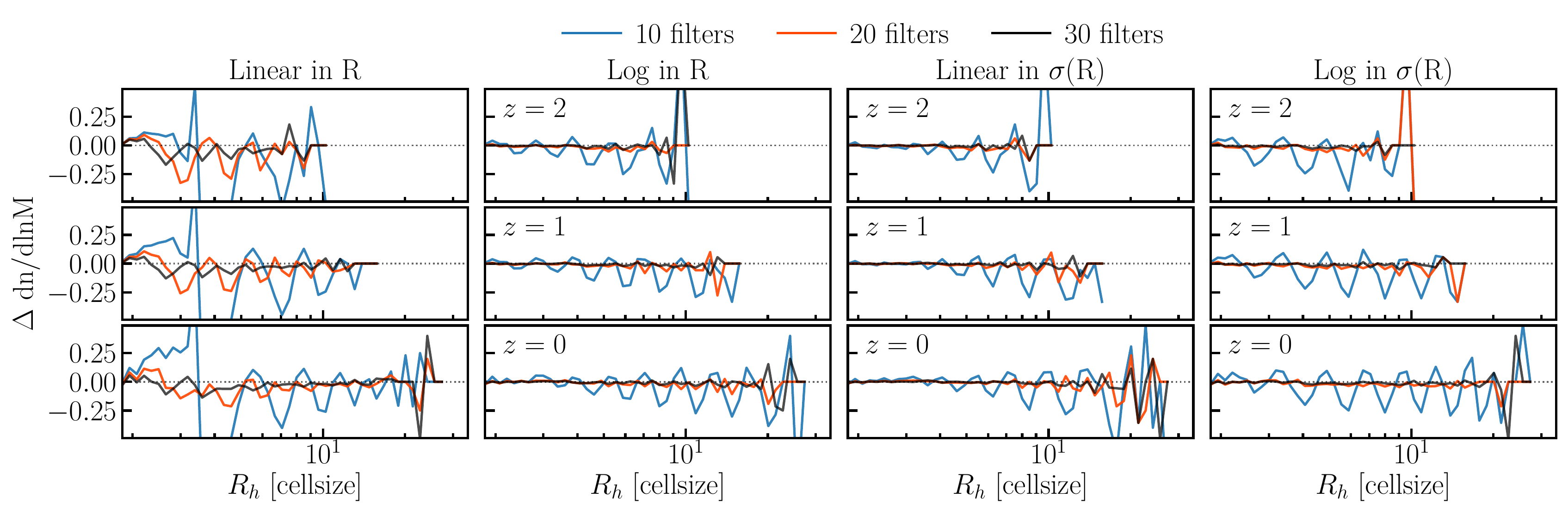}
\end{center}
\vspace{-0.3cm}
\caption{To maximize the computational performance of the mass-Peak Patch method we converged on an optimal `filter bank' by running a suite of 1024$^3$ particle, 1024 Mpc simulations at $z=[0,1,2]$, which varied the number and spacing of filters. Each column (type of filter spacing) shows the convergence of the mass-Peak Patch mass function when varying the number of filters, referenced to the `converged' case of 100 filters. From left to right we present linear spacing in $R_f$, logarithmic spacing in $R_f$, linear spacing in $\sigma(R_f)$, and logarithmic spacing in $\sigma(R_f)$. Each row shows a different redshift, from $z=2$ at the top to $z=0$ at the bottom. Colours represent the number of filters used in each filter bank, where blue, red, and black indicate 10, 20, and 30 filters, respectively. Filters are spaced from a minimum size of R$_{\rm{min}}$=2Mpc to a maximum size of R$_{\rm{max}}$=36Mpc.}
\label{fig:filters}
\end{figure*}

\subsection{Optimal Mass-Peak Patch Parameters}
\label{sec:peakpatchparameters}

Though we do not vary the free parameters of the mass-Peak Patch model with cosmology or resolution, and no parameters are fit to N-body simulations, there still remain a few definitions of halo collapse and exclusion parameters that need to be chosen in order to perform a simulation, and to maximize computational efficiency. These include the overdensity at which a halo is considered to have collapsed, how to divide overlapping volumes of Lagrangian space belonging to separate halo candidates in order to avoid double counting of mass, and how to efficiently find the locations of interest at which to calculate the ellipsoidal collapse equations. These choices are discussed and validated in the remainder of this section, and the optimal parameter findings settled upon are listed in Table~\ref{tab:choices}. 

\subsubsection{Density-Peak Finding to Generate Candidate Mass-Peak Points} 
\label{sec:peakfinding}

The mass-Peak Patch method is a halo finder that identifies halos with the largest regions of Lagrangian initial-state space that will collapse by the redshift in question using approximate dynamics, rather than the more exact N-body dynamics used for final-state Eulerian halos. The first mPP step after the generation of the initial conditions is to generate a point process of possible sites of halo collapse. A highly overcomplete set is to use every initial-space lattice point, adjusting the Lagrangian radius of the region until the nonlinear ellipsoid trajectories, homogeneous over that radius in all three directions, would have compactified to near-zero volume. In its fully glory this is the correlated excursion set approach. Afterwards, the final complete set of non-overlapping halos would be found by an exclusion process. 

In practice, such dense sampling of candidates is a computationally-expensive overkill. BM used a much smaller subset of linear density peaks smoothed on a hierarchy of filter scales $\{ R_f \}$, chosen to lie above a linear density threshold $\delta_c$ low enough that it would not cause collapsed structures to be missed. The concrete choice $\delta_c\sim1.5$ was made, safely below the $\delta_{\rm{crit}} = 1.686$ of top hat spherical overdensity collapse, and the catalogues were shown to be insensitive to the specific value chosen. We continue with that choice in our modernized version. Smoothing is performed by convolving the density field with a top hat filter of the form $W(kR_f)=3j_1(\kk R_f)/(\kk R_f)$, where $j_1$ is the spherical Bessel function of order 1. This differs from the BM implementation in which Gaussian smoothing was used, since we found that this gives  more candidate peak-points on a given scale than Gaussian or sharp $k$-space filtering, and is better matched to how we actually average various fields about the candidate points to determine the ellipsoidal dynamics, namely with top-hat filtering over scales $R_c$. The number density of peaks on a coarse-grained scale $R_f$ scales as $R_f^{-3}$, meaning the smallest filters in the hierarchy dominate the candidate peak list, so there is much room for modifying the distribution of filters without adding much more expense than that of the smallest element in the multi-resolution hierarchy. 

Thus, in order to optimize performance, we want to determine the best number of filters and their spacing for our `filter bank'. Too few filters may cause potential halos in the simulation to be missed, while too many filters, especially of small scale, will reduce the computational performance of the code. The filters need to span a range that covers the mass range of halos that we expect to find. The largest halo expected to be found at redshift $z=0$ has a mass of $\sim$$10^{16}M_\odot$, corresponding to a radius of $\sim$36Mpc in Lagrangian space, while the smallest halo is set by the resolution of the simulation. At a larger redshift one can use a universal halo mass function to quickly estimate the size of the largest halo expected to be found (e.g. \citet{2008ApJ...688..709T}). For the homogeneous ellipsoidal calculations we use spherical averaging, which essentially imposes a minimum number of lattice sites needed for good smoothed-field resolution. Here we set the minimum radius to be $2a_{\rm{latt}}$, two lattice spacings of size $a_{\rm{latt}}$ in the simulation (also called the cell size in this work). This therefore defines a minimum halo mass that we can possibly resolve: about each candidate point at least 32 neighbours are needed to encompass the top-hat smoothing estimates, similar to the minimum count used in smoothed particle hydrodynamic simulations. For more accurate sphericalization, a minimum count of 100 lattice ``particles'' is more conservative, and is the typical number we use to define our minimum halo target mass in the simulations. The speed of the mass-Peak Patch method means that there is no real problem with making larger and larger lattices, so we can design a simulation to achieve a desired minimum resolved halo mass given a specified box size.     

With the physical arguments setting the minimum and maximum filter sizes to use, we have the freedom to choose how many filters lie within in this range, and what their spacing is. Logarithmic spacing in $R_f$ is what BM used. Linear in $R_f$ is another possible choice, or some combination of logarithmic on large scales and linear on small to control the low $R_f$ regime. The number of filters to well-sample candidate mass-peaks is determined empirically. In this work we also investigated using filters spaced by the field variance on the scale of the filter,
\begin{equation}
\sigma^2(R_f) = \frac{1}{(2\pi)^3}\int k^2 P(k) W^2(kR_f) dk,
\end{equation}
which was not done in the original Peak Patch implementation, but is logical to consider since the mass functions are dominantly functions of $\sigma(R_f)$ \citep{1991ApJ...379..440B, 2008ApJ...688..709T}. To converge on an optimal choice we ran a suite of 1024$^3$ particle, 1024 Mpc mPP simulations at $z=[0,1,2]$ which varied the filter spacing between linear and logarithmic in $R_f$ and in $\sigma(R_f)$, and using a total number of filters ranging from 5 to 100. The effects of the filter choice on the halo mass function of the resulting mPP simulation, relative to the `converged' case of 100 filters, can be seen in Figure~\ref{fig:filters}. We chose to plot results for 10, 20, and 30 filters, as using greater than 30 was nearly indistinguishable from 30, and using less than 10 was clearly insufficient. It is evident that using linearly spaced filters in $R$ (first column) is inferior to using log spaced filters in $R$ (second column). The linear spaced filters leave large gaps and spikes in the mass function, especially at the low mass end, where collapsed regions in the simulation have not been identified, and therefore the halos they correspond to were not found. The logarithmically spaced filters do much better for the same total number of filters (same computational cost). For filter spacing in $\sigma(R)$ both linear and logarithmic spacing are roughly equivalent, and give similar results to filters logarithmically spaced in $R$. It is also clear that more than 10 filters are needed, as each spike seen when using only 10 filters corresponds to a filter, and halos in between these masses were missed. We find that between 20 and 30 filters are sufficient. 

This convergence test was conducted at a single simulation resolution, so the total number of filters is not the most robust way to create the filter bank - the filter \textit{spacing} is. As a rule of thumb, the number of filters converged upon above (20-30) correspond to a filter spacing of: $\rm{\Delta ln R}$=0.15 to 0.1, $\rm{\Delta \sigma(R)}$=-0.08 to -0.06, or $\rm{\Delta \sigma(R)}$=-0.09 to -0.06. These filters should be spaced from a minimum radius of R$_{\rm{f,min}}$ = 2a$_{\rm{latt}}$ to a maximum radius of the largest halo expected to be found at the redshift in question (R$_{\rm{f,max}} \sim $36Mpc at z=0).

\subsubsection{Homogeneous Ellipsoid Collapse} 
\label{sec:ellipsoidalcollapse}

\begin{figure*}
\begin{center}
\includegraphics[width=0.71\textwidth,trim={0 0 5 0},clip]{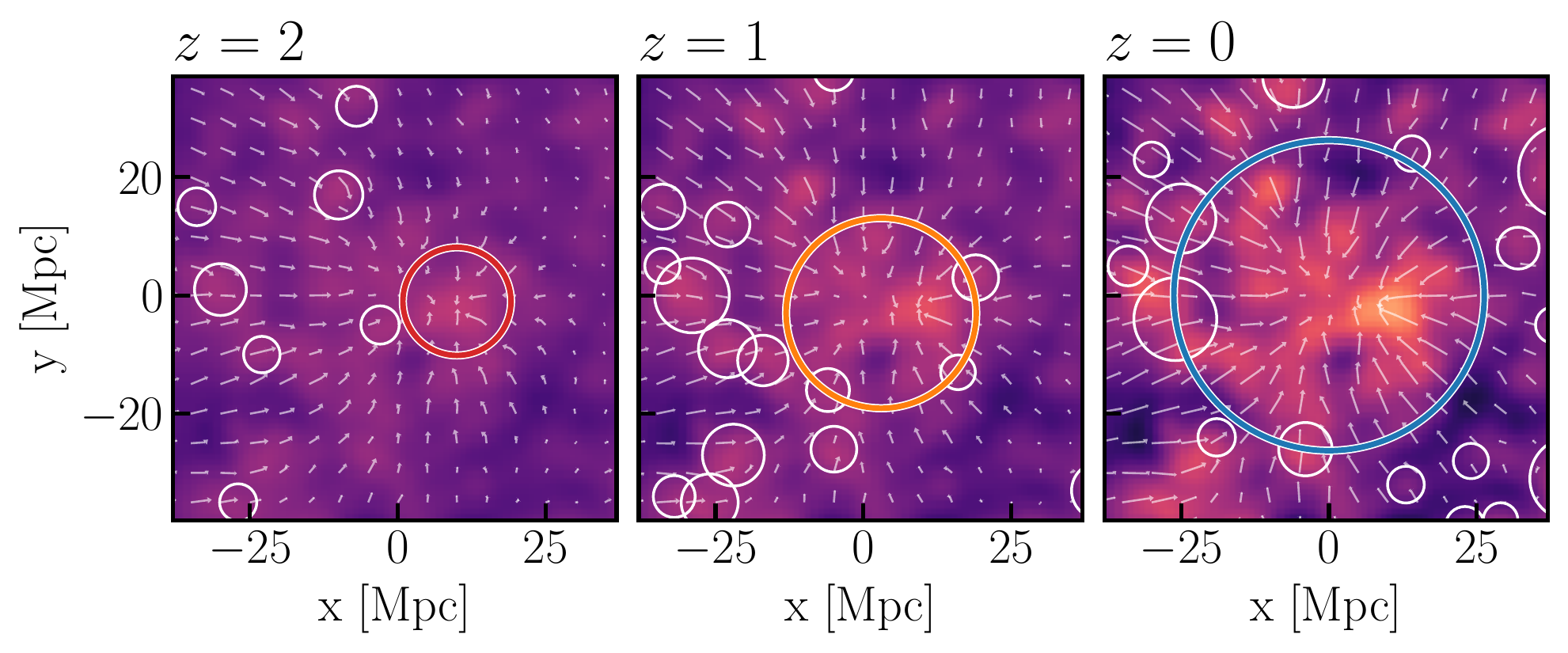}
\includegraphics[width=0.28\textwidth,trim={7 -8 0 0},clip]{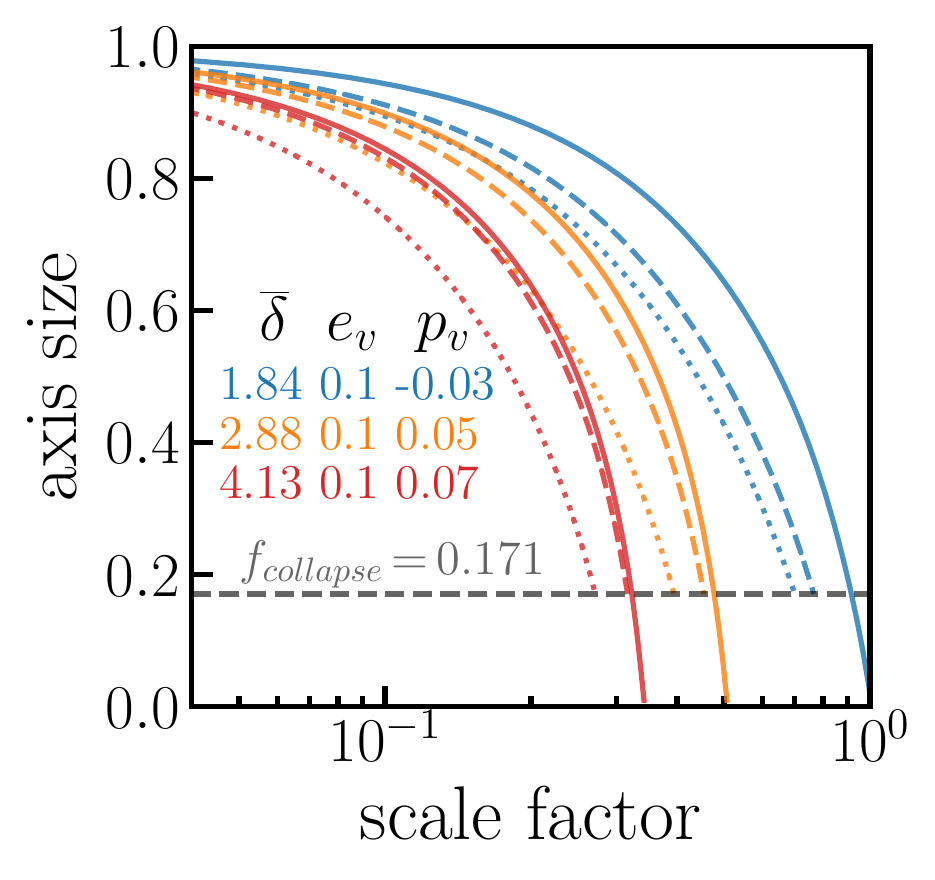}
\end{center}
\vspace{-0.5cm}
\caption{{\bf{left three panels:}} Evolution of halos in a 70 comoving Mpc region at three redshifts, where circles represent the radius of the Lagrangian halos determined by the mass-Peak Patch method, and the background image shows the linear density field scaled by the linear growth factor. The growth and merging of halos is immediately apparent as the central halo (coloured) grows from a mass of $1.7 \times 10^{12}M_\odot$ at $z=2$, to a mass of $2.6 \times 10^{15}M_\odot$ at $z=0$. White arrows show the Lagrangian displacement field, scaled down by a factor of 5, in the reference frame of the central halo at each redshift. Arrows are bent as a result of plotting the 1LPT (arrow tail) and 2LPT (arrow head) components of the displacement separately. {\bf{right panel:}} Evolution of the homogeneous ellipsoid axes (co-moving size) for the most massive halo at each redshift shown on the left. Dotted, dashed, and solid lines show the collapse along the third, second, and first axis of the ellipsoid, respectively, while the colours (matched with the 3 left panels) indicate the redshift. The third and second axes are stopped at a radial freezout factor of $\Delta_h^{-1/3}=0.171$, and the complete collapse of the first axis marks the formation redshift of the halo. Included are the values of the halo's mean density $\overline{\delta}$, ellipticity $e_\nu$, and prolaticity $p_\nu$, which completely determine it's collapse as per equation~\ref{eq:ellipsoid}.}
\label{fig:singlehalo}
\end{figure*}

The final size of a mass-Peak Patch halo is determined through a set of homogeneous ellipsoid collapse equations, which take as input the linear strain $\eonebar_{ij}(R_{p_1})\equiv (\partial{\sonebar_i}/\partial{q_j}+\partial{\sonebar_j}/\partial{q_i})/2$ averaged over a sphere of Lagrangian radius $R_{p_1}$, centered on each halo candidate. The mean strain within a given measurement radius is diagonalized to obtain its eigenvalues, $\lambda_i=-(\overline{\delta}\!/3)(1+c_i)$, and eigenvectors, $\nhat_i$, related through $\ebar_{ij}=\sum_k \lambda_k\nhat_k^i\nhat_k^j$, where
 $c_3=p_\nu+3e_\nu$, $c_2=-2p_\nu$, and $c_1=p_\nu-3e_\nu$. The redshift at which the corresponding homogeneous ellipsoid would collapse is calculated from the following system of ordinary differential equations:
 
\begin{equation}
\frac{\ddot{x}_i}{x_i}= \frac{\ddot{a}}{a} - \frac{1}{2}\Omega_mH^2\left[b_i\dbar+c_i\dbar_{\rm lin}\right].
\label{eq:ellipsoid}
\end{equation}
$a$ is the background scale factor, $x_i(t)=R_i(t)/R_{m_1}$ are the scale factors in each of the three principal axes of the ellipsoid, $\dbar=a^3/(x_1x_2x_3)-1$, $\dbar_{\rm lin}=\dbar{D}$, and 
\begin{equation}
b_i(t)\equiv  \frac{3}{2}\int_0^\infty d\tau [\tilde{x}_i^2 +\tau]^{-1}\prod_j [\tilde{x}^2_j+\tau]^{-1/2}.
\end{equation}

The parameters $b_i$ are defined by the Newtonian potential of an isolated homogeneous ellipsoid in a coordinate system aligned with its principal axes: $\Phi(\x)=4\pi{G}\rho\sum_i b_ix_i^2/2$, and depend only on its evolving shape. Thus, the first term in the brackets on the right hand side of equation (\ref{eq:ellipsoid}) is proportional to the gravitational acceleration from material inside of the ellipsoid, and is exact, while the second term represents that from external tidal fields, and is approximated by the linear solution, which only depends on the mean linear strain within $\rbar$, through $c_i$ \citep{1996ApJS..103....1B}. The parameters $b_i$ and $c_i$ can be considered to encode the anisotropy of the internal and external forces, respectively. Since the shape of the ellipsoid is time dependent, while the shape of the linear tidal field is not, $b_i$ is time dependent while $c_i$ is not. Note that in the spherical case, $x_1=x_2=x_3=x$ implies $b_i=1$, $c_i=0$, and $\ddot{x}/x = \ddot{a}/a - \Omega_mH^2\dbar/2$, which is the usual equation of motion for spherical collapse.

Each of the three axes of the ellipsoid is evolved until it reaches a critical collapse radius of $x_{\rm{coll},i} = a f_{\rm{coll},i}$, after which the axis is `frozen in' at that value, as seen in the right panel of Figure~\ref{fig:singlehalo}. The eigenvalues of the ellipsoid satisfy $\lambda_3 \geq \lambda_2 \geq \lambda_1$, so the 3-axis will be the first to collapse, and the 1-axis will be the last. In the mass-Peak Patch picture the collapse of the 1-axis marks the virialization of the halo into a collapsed object. If a halo of radius $R_{p_1}$ has collapsed by the redshift in question, we then perform the measurement at increasingly larger radii, $R_{p_n}$ > ... > $R_{p_1}$, until the halo centered at that location no longer collapses. If the halo has not collapsed at $R_{p_1}$, we step down to increasingly smaller radii. The largest region found to collapse at that location is then appended to the initial halo catalogue. 

There remains freedom for the initial search radius $R_{p_1}(R_f)$, where $R_f$ is the filter scale that the density peak was found at, and freedom for the choices of the three radial freezeout factors, $f_{\rm{coll},3}$, $f_{\rm{coll},2}$, and $f_{\rm{coll},1}$. The initial search radius is a parameter that can help to minimize under-prediction of halo sizes while maximizing computational efficiency. In theory, to ensure the largest halos are always found, one would choose $R_{p_1}$ to be the size of the largest halo expected to be found in the entire simulation, $R_{p_1}=R_{h,max}$, and then step down to increasingly smaller radii until the first radius at which a halo collapses. But, this can be very inefficient, as the cost of calculating the mean strain used in the collapse equations scales as $R^3$ (for a 0.5 Mpc resolution simulation one would then be summing over $\sim$1$\times$10$^6$ cells for every halo). Conversely, if $R_{p_1} \simeq R_f$, while stepping out to larger radii, a larger halo (for example of size $1.3R_f$) would not be identified even though it would collapse, if a smaller halo (for example $1.2R_f$) does not collapse, as the calculation is stopped when the halo centered at that location no longer collapses. We found that choosing $R_{p_1} = 1.75 R_f$ results in a nearly identical halo catalogue to that of the theoretical best choice of $R_{p_1}=R_{h,max}$, at a fraction of the computational cost. This choice was adopted as the default for all future simulations.

The radial freeze-out factors $f_{\rm{coll},i}$ are fundamental for the homogeneous ellipsoid approximation to gravitational collapse.
Two physically motivated values are $f_{\rm{coll},i}=\Delta_h^{-1/3}$, which corresponds to the standard top hat virial density contrast of $\Delta_h$, and $f_{\rm{coll},i} \sim 0$, which corresponds to complete collapse along the axis. For this work we have adopted the halo definition of $\Delta_h = 200\overline{\rho}_m$, where $\overline{\rho}_m$ is the mean density of the universe. Therefore, we have the choice of $f_{\rm{coll},i}=200^{-1/3}=0.171$, or $f_{\rm{coll},i} \sim 0$. 

\begin{table*}
\begin{center}
\begin{tabular}{l l c }
Algorithm Step & Required Parameters & Finalized Choices\\ 
\hline
\hline
Density-Peak Finding~(\ref{sec:peakfinding}) & Type of filters & Top hat \\ 
  &  Spacing between filters & Logarithmic in R or $\sigma(R)$, or Linear in $\sigma(R)$ \\ 
 &  Total number of filters & 20-30 \\ 
\hline
Ellipsoidal Collapse~(\ref{sec:ellipsoidalcollapse}) &  Initial search radius & 1.75 R$_{f}$\\ 
 & Overdensity of a halo$^\dagger$ & $\Delta_h$ = 200 $\overline{\rho}_M$\\ 
 & Axis `freeze out' ratio$^\dagger$ & ($\lambda_3,\lambda_2,\lambda_1$)=($\Delta_h^{-1/3}$, $\Delta_h^{-1/3}$, 0.01)\\ 
\hline
Exclusion~(\ref{sec:exclusion}) & Partition of overlapping Lagrangian space$^\dagger$ & Binary exclusion, type (ii) \\ 
\hline
Displacements & First or second order LPT & Second order\\ 
\end{tabular}
\caption{Parameters of the mass-Peak Patch method that need to be specified in advance in order to run a simulation. Entries followed by a $^\dagger$ are fundamental to the physical choice of approximating gravitational collapse used in the mass-Peak Patch method, while those without a $^\dagger$ are simply to maximize computational efficiency.} 
\label{tab:choices}
\end{center}
\end{table*}

To determine the effect of axis freeze-out on the final halo catalogue we ran a suite of simulations with the following four axis choices:
\begin{align}
\begin{split}
(f_{\rm{coll},3},f_{\rm{coll},2},f_{\rm{coll},1}) &= (\Delta_h^{-1/3},\Delta_h^{-1/3},\Delta_h^{-1/3}) \\  
&= (\Delta_h^{-1/3},\Delta_h^{-1/3},0.01) \\
&= (\Delta_h^{-1/3},0.01,0.01) \\
&= (0.01,0.01,0.01).
\end{split}
\end{align}
A value of 0.01 was chosen for complete collapse along an axis as it was near enough to complete collapse, but not sufficiently small to cause any unwanted numerical effects from calculating the differential equations. We found that requiring complete axis collapse along the final axis, and virialization along the first 2, i.e. $(f_{\rm{coll},3},f_{\rm{coll},2},f_{\rm{coll},1})$=$(0.171,0.171,0.01)$, results in halo catalogues that most closely match common universal halo mass functions, so we adopt this for our runs. The other three choices over-predicted the number of halos at the high mass end.

\subsubsection{Exclusion} 
\label{sec:exclusion}

\begin{figure}
\begin{center}
\includegraphics[width=.9\columnwidth,trim={0 0 0 0},clip]{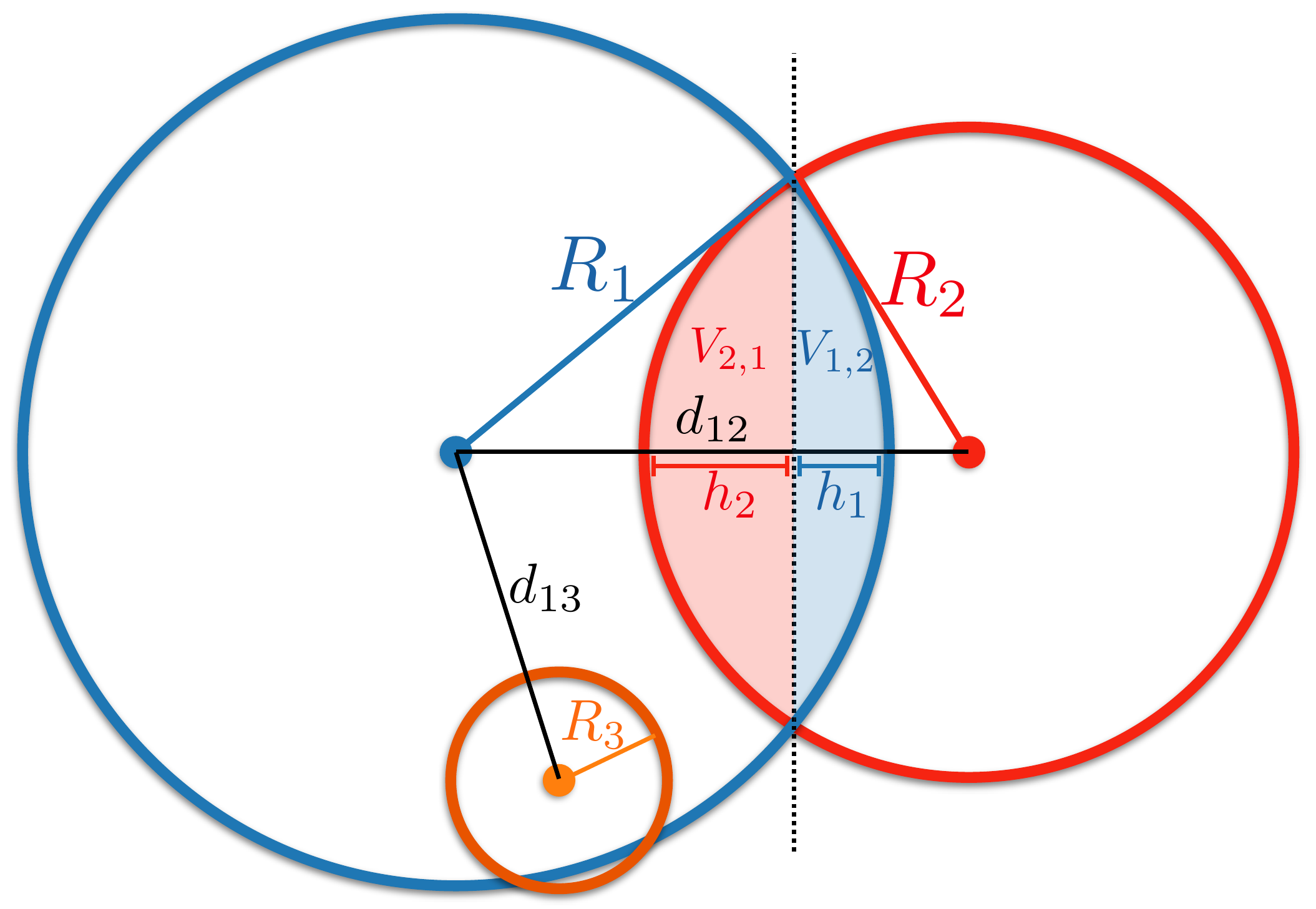}
\end{center}
\vspace{-0.5cm}
\caption{Overlapping halos in Lagrangian space are dealt with through exclusion and reduction. Considering the binary merge (type (ii)) used for the N-body comparisons of Section~\ref{sec:comparisons}, patch $3$ would be removed from the list of candidate halos, and both patches $1$ and $2$ will see a reduction in volume (mass) of $V_1 = V_1 - V_{1,2}$ and $V_2 = V_2 - V_{2,1}$, respectively. Although the volume reduction occurs only as a consequence of the spherical caps of the union of the two spheres, we treat it as a reduction of halo radius which retains the sphericity of the halo.}
\label{fig:exclusion}
\end{figure}

The final choice in the mass-Peak Patch method is how to ensure that collapsed regions are distinct and that there is no double counting of matter. After the computation of the homogeneous ellipsoid collapse there will be candidate peak patches that overlap, and there will be smaller collapsed objects enveloped in larger collapsed patches. In order to prevent this unphysical result propagating to the final halo catalogue we must perform a hierarchical Lagrangian exclusion and reduction algorithm.

One may think of a number of ways of doing this, with many discussed in detail in \citet{1996ApJS..103....1B}. Figure~\ref{fig:exclusion} depicts the geometrical problem that needs to be solved in the simple case of 3 overlapping halos. Two options are \textit{binary exclusion} and \textit{half exclusion}. In both of these methods, candidate mass-peak patches are first ranked in mass, and for each patch all neighbors that have any overlap with its Lagrangian radius are considered. If the center of a smaller patch resides inside the Lagrangian volume of a larger patch, the smaller patch is removed from the list (excluded). For the half exclusion option this is the only step, and the remaining halo catalogue is then written to disk. For binary exclusion, another step, called \textit{reduction}, is implemented in order to prevent any double counting of Lagrangian mass. If adjacent patches have any overlap we calculate the volume (mass) of the union of the two intersecting spheres by the standard equations of the volume of a spherical cap, 
\begin{align}
V_{i,j} &= \frac{1}{3} \pi h_i^2 (3R_i-h_i) .
\end{align}
The height of the cap $h_i$ overlapping sphere $j$ can be written as
\begin{align}
h_i &= \frac{(R_j - R_i + d_{ij})(R_j + R_i - d_{ij}) }{2d_{ij}}
\end{align}
where $R_i$ is the radius of sphere $i$ .

Considering the configuration shown in Figure~\ref{fig:exclusion}, we have a patch $p_3$ that will be excluded, and 2 overlapping spheres $p_1$ and $p_2$, where the volumes of overlap are  $V_{1,2}$ and $V_{2,1}$. There are then 3 options of how to partition this overlapping mass: 
\begin{enumerate}
\item Subtract the total mass ($V_{1,2}$+$V_{2,1}$) from the larger patch.
\item Consider the plane perpendicular to the vector ${\bf{d}}_{ij}$ which divides the two spherical caps. Subtract any volume beyond this plane from each halo, $V_1 = V_1 - V_{1,2}$,  $V_2 = V_2 - V_{2,1}$. 
\item Subtract the total mass ($V_{1,2}$+$V_{2,1}$) from the smaller patch.
\end{enumerate}

The overlapping mass from all neighboring halos is accounted for before reducing the volumes. Although the mass reduction occurs only on the overlap between peaks, we treat it as a reduction in radius which retains the sphericity of the patch. These reductions conserve mass, but as the patches are approximated as spheres, there may still be a small overlap in volume after, which we ignore. The final radius of a halo is easily related to the mass through $M_{\rm h} = 4\pi R_{\rm h}^3\overline{\rho}_{M}/3$, where $\rm{\bar\rho_M = 2.775 \times 10^{11} \Omega_M h^2\ [M_\odot/Mpc^3]}$.

The effects of binary exclusion, compared with half exclusion, were studied through a series of simulations at varying redshifts and resolutions. As expected we found that half exclusion has more halos at all masses, as a result of not conserving mass of overlapping volumes. We find that just as in the original implementation of the mass-Peak Patch method, binary exclusion of type (ii) is currently the best choice. There remains the possibility to design and implement a more physical method to perform exclusion, but this is left for future work. 

We note that although there most likely exists an optimal combination of choices for the filter bank, axis-freezeout, and exclusion, that might maximize a specific summary statistic when compared to a specific N-body, we do not attempt to do this. Overall, we found that the BM choices have stood well against the test of time. The parameter choices we described in this section can be seen in Table~\ref{tab:choices}.

\section{mass-Peak Patch N-body Validation}
\label{sec:comparisons}

In this section we validate the mass-Peak Patch method against N-body simulations through a number of summary statistics. By matching the initial conditions of a mPP simulation to those of the N-body run we are able to compare halo mass functions, halo power spectra \& cross correlations, and higher order spatial matching.

\subsection{Reference Simulations}
\begin{table*}
\begin{center}
\begin{tabular}{c c c c c c c c}
Method & N$_{\mathrm{runs}}$ & N$_{\mathrm{cells}}$ & L$_{\mathrm{box}}$ [Mpc] & Redshift & N$_{\mathrm{processors}}$ & Total Runtime [s/run] & Memory [Gb]\\ 
\hline
Peak Patch & 56 & 1024$^3$ & 1024 & 0,1,2 & 32 & 185 & 35 \\[0.2cm] 
N-Body & 3 & 512$^3$ & 1024 & 0,1,2 & 200 & $\sim$168,000 & 61 \\ 
Peak Patch & 3 & 512$^3$ & 1024 & 0,1,2 & 8 & 100 & 15 \\[0.2cm] 
N-Body & 3 & 512$^3$ & 512 & 0,1,2 & 200 & $\sim$168,000 & 61 \\ 
Peak Patch & 3 & 512$^3$ & 512 & 0,1,2 & 8 & 140 & 15 \\[0.2cm] 
N-Body & 3 & 512$^3$ & 256 & 0,1,2 & 200 & $\sim$168,000 & 61 \\ 
Peak Patch & 3 & 512$^3$ & 256 & 0,1,2 & 8 & 300 & 15 \\ 
\end{tabular}
\caption{A list of all validation runs included in this paper. The compute time to generate the initial conditions is not included in the timing, as this was done externally, but the timing does include reading the initial conditions from disk. Note that although the speed-up factor of these small $512^3$ validation runs is generally greater than a factor of $10^3$, this factor becomes increasingly large for more massive runs, due to the straightforward parallelization of the mass-Peak Patch method described in Appendix~\ref{sec:appendixa}. We found a speed up factor of $>3000$ when comparing to the 4096$^3$ particle Gadget-2 Mice Grand Challenge \citep{2015MNRAS.448.2987F,2015MNRAS.447..646C,2015MNRAS.447.1319F}, the results of which will be presented in a future work.}
\label{tab:runs}
\end{center}
\end{table*}

To test the accuracy of the mass-Peak Patch method we created a suite of N-body simulations using the publicly available Gadget-2\footnote{https://wwwmpa.mpa-garching.mpg.de/gadget/} code. \citep{2005MNRAS.364.1105S}. We ran a set of 512$^3$ particle simulations with box-sizes equal to 256 Mpc, 512 Mpc, and 1024 Mpc. Throughout this paper we refer to these three sets of simulations by the comoving resolution of the 3-D grid of the initial conditions: the \textit{0.5 Mpc resolution} simulation, the \textit{1 Mpc resolution} simulation, and the \textit{2 Mpc resolution} simulation. For each resolution we ran three independent realizations, which we refer to as different simulation \textit{seeds}, resulting in a suite of 9 simulations in total. The following cosmological parameters were used, which match those of the Euclid comparison project: $\Omega_M=0.285$, $\Omega_b = 0.044$, $\Omega_\Lambda = 0.715$, $h=0.695$, $n_s=0.9636$, and $\sigma_8=0.828$.

The initial conditions were generated with the publicly available 2LPTic code\footnote{http://cosmo.nyu.edu/roman/2LPT/} using a starting time of $a=0.02$ ($z$=49) and a linear power spectrum generated with CAMB\footnote{https://camb.info/}. The comoving force softening length ($\epsilon$ = 20h$^{-1}$kpc), absolute force error tolerance ($\alpha$ = 0.002), and time stepping ($\Delta$ ln($a$)$_{\rm{max}}$= 0.0125) chosen was informed by the convergence tests of the Aemulus project \citep{2018arXiv180405865D}. If we were instead targeting validations at high redshift regime the starting redshift would be increased to $z\sim100$. 

To perform halo finding we used the Adaptive Mesh Investigations of Galaxy Assembly (AMIGA) halo finder\footnote{http://popia.ft.uam.es/AHF/Download.html}, also known as AHF. We defined dark matter halos as Eulerian-space spherical overdensities - which are themselves mass-peak patches - of 200 times the background matter density. We use `strict' spherical overdensity masses, where strict refers to the inclusion of unbound particles residing within a halo. This choice was made since we are mainly interested in the total mass of the halo and not its internal properties for this study.

For our mass-Peak Patch runs we modified 2LPTic to output the density and displacement fields, linearly evolved to redshift 0, instead of the 2LPT displaced particles. Each mass-Peak Patch simulation used the internal parameters specified in Table~\ref{tab:choices}. 

To create the reference Lagrangian N-body halo catalogues we traced back each N-body halo's constituent particles to their initial positions and calculated their centre of mass, which we define as the halo's Lagrangian position. A halo's Lagrangian radius is then simply $\rm{R_{h} = (3M_{h}/4\pi\overline\rho_{M}))^{1/3}}$.

A list of all runs used in this paper and their computational requirements are shown in Table~\ref{tab:runs}. 

\begin{figure*}
\begin{center}
\includegraphics[width=1.\textwidth,trim={0 0 0 0},clip]{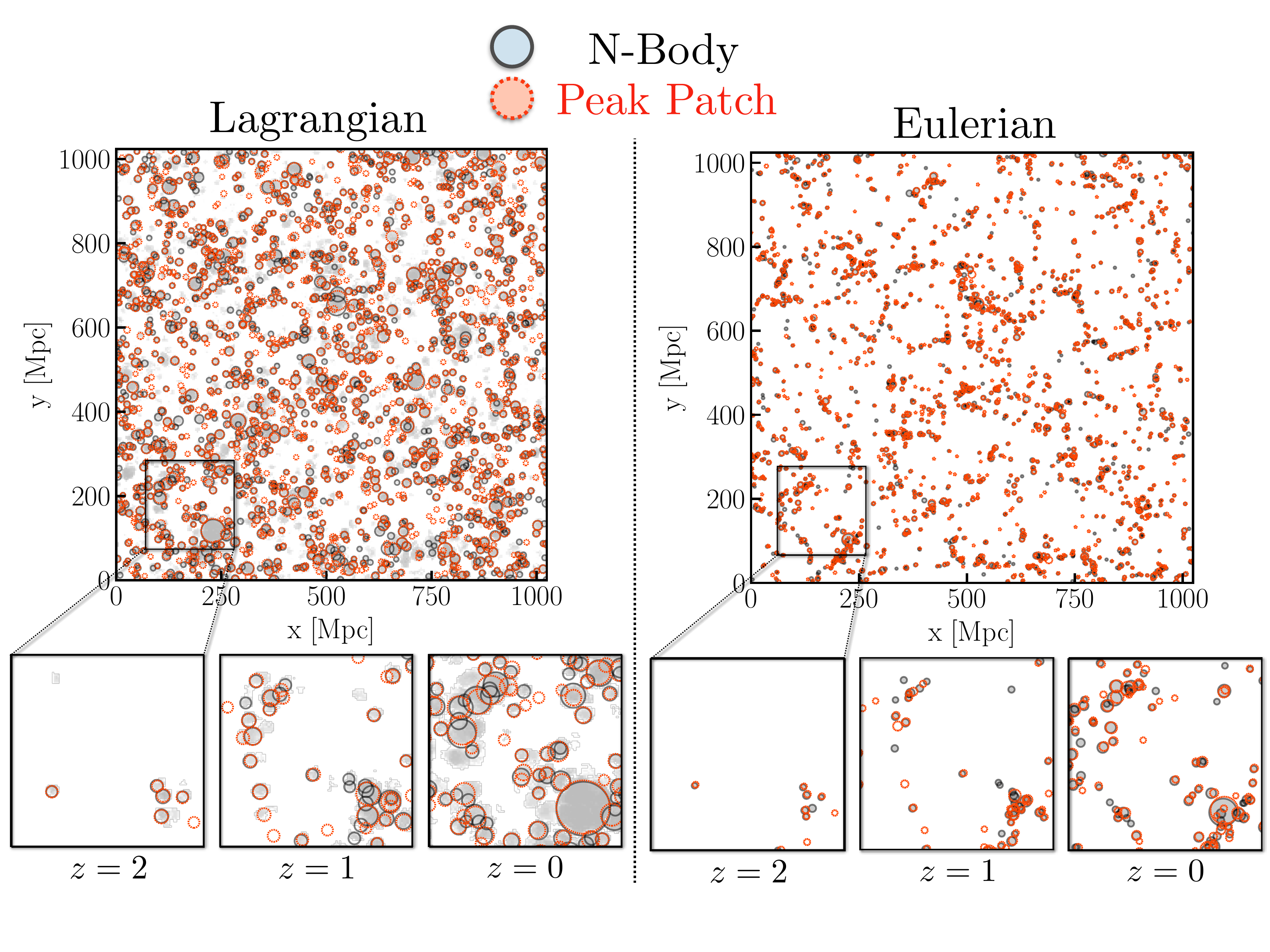}
\end{center}
\vspace{-0.3cm}
\caption{Lagrangian (left) and Eulerian (right) halos of a 1024 Mpc simulation for both N-body (black, solid lines) and mass-Peak Patch (red, dashed lines). The top panels show the full simulation volume at $z=0$, while the bottom show a 200 Mpc zoom-in at $z=[2,1,0]$ centered near the largest halo at $z=0$. Lagrangian N-body halo positions were created by finding all particles belonging to each Eulerian halo, tracing the particles back to their initial positions (depicted here by the diffuse black background), and calculating their center of mass. Visualized here is a 20 Mpc thick slice in the $z$-direction. The radius of the Lagrangian halos is proportional to their mass by $R_h=(\frac{3}{4\pi \overline{\rho}_M}M)^{1/3}$, while the Eulerian radius is a factor of three larger than $R_{200,M}$ for visualization purposes.}
\label{fig:halos}
\end{figure*}

\subsubsection{Halo Clustering Benchmarks}
\label{sec:benchmarks}

As in \citet{1996ApJS..103....1B}, one of the most compelling ways to compare the mass-Peak Patch method to N-body is by direct visual comparison of the positions of the halos in the two catalogues, since this involves high order clustering as well as the traditional low order two-point correlations. To make that more precise mathematically we introduce a new high order statistic, by removing the mPP halos residing inside of N-body halos, and vice versa, using our exclusion/merger post-processing. We entitle this a \textit{halo spatial matching} comparison, and present the results in Section~\ref{sec:lag_partner}.

We also use more traditional 2-point statistics in Section~\ref{sec:clustering} to show the power spectra of the catalogues are quite similar. We quantify this by a near-unity halo transfer function $T(k)$, and by a more demanding halo cross correlation coefficient, $r(k)$, which probes the mismatch of halo centres in the 2-catalogues. Ideally this would have a halo-mass-dependent scaling to avoid small object and large object offsets being treated exactly the same. We have not done this scaling here, since, in many respects, the first test of merging of one catalogue relative through exclusion is the more demanding scaling test.   

For the 2-point tests, the halo power spectra are calculated  by first binning halos onto a cubic grid to create the halo density field, $\delta_{h}(\x)$, or its Fourier transform $\delta_{h}(\kk)$, where $h$ represents halos from the N-body or mass-Peak Patch methods: $h=[NB, PP]$. The power spectrum is then defined as $\langle \delta_{h}(\kk)\delta_{h^\ast}(\kk')\rangle = (2 \pi)^3 \delta_D^3(\kk+\kk') P_{h,h^\ast}(k)$, which we can simply calculate for each linearly spaced bin $k_i$ through 

\begin{align}
\begin{split}
P_{h,h^\ast}(k_i) &= \int_{|\kk| \in k_i} \frac{d^3\kk}{V_{k_i}} \delta_h({\bf{k}})\delta_{h^\ast}(-{\bf{k}}) \\
&= \frac{1}{V_{cell}} \sum_{|\kk| \in k_i} \frac{1}{n_{k_i}} \delta_h(\kk)\delta_{h^\ast}(-\kk) .
\label{eq:powerspectrum}
\end{split}
\end{align}
The second equality holds when the calculation is performed on a discrete periodic grid as in our analysis. $V_{\mathrm{cell}}$ is the volume of a cell, and $V_{k_i} \approx 4\pi k_i^2 \Delta k$. We used the python package \texttt{nbodykit} \citep{2017arXiv171205834H} to perform some of these calculations. 

$P_{h,h^\ast}(k)$ allows us to define the transfer function $T(k)$ as the ratio of the square root of the power spectra,
\begin{align}
T(k) &= \sqrt{\frac{P_{PP,PP}(k)}{P_{NB,NB}(k)} }\ ,
\end{align}

and the cross correlation coefficient, $r(k)$, as 
\begin{align}
r(k) &= \frac{P_{NB,PP}(k)}{\sqrt{P_{NB,NB}(k) P_{PP,PP}(k)}}\ .
\end{align}

The transfer function is a measure of the relative bias between halos. The cross correlation coefficient is related to another measure of bias: if the two halo number-density fields, $n_{PP}(x)$ and $n_{NB}(x)$  are considered to be in the correlated Gaussian random field limit, appropriate for low $k$, then the mean response of the mPP density to a given N-body density is 

\begin{align}
\begin{split}
<\tilde{n}_{PP}(k) \vert \tilde{n}_{NB}(k)> &= \frac{P_{PP,NB} (k)} {P_{NB,NB} (k)}\  \tilde{n}_{NB}(k)\\ &=  \frac{r(k)}{T(k)}\ \tilde{n}_{NB}(k). 
\end{split}
\end{align}
As mentioned above it would be better to band halos of nearly the same mass together, and do the cross-test for the individual mass-bands, each scaled with a radius $\propto M^{1/3}$, but here we shall be content with just showing the global results, with one very coarse halo-mass band. All results we show in this work are calculated after abundance-matching the mass-Peak Patch halo catalogues to those of the N-body, which is accomplished by rank ordering in halo mass and selecting the same number of the most massive halos for each. The labels shown on the figures then represent the number of particles or mass of the N-body halos.

\subsection{Halo Visuals}
\label{sec:halo_visual}

Figure~\ref{fig:halos} provides a first, qualitative comparison of mass-Peak Patch halos with those of an N-body simulation in both Lagrangian and Eulerian space. The large panels on top show a 20 comoving Mpc thick slice of the full 1024 Mpc simulation at $z=0$ in Lagrangian space (left) and Eulerian space (right), while the zoom-ins on the bottom show a 200 Mpc region at $z=[2,1,0]$ centered near the largest halo in the simulation at $z=0$. Black circles represent the Lagrangian or Eulerian radius of the N-body halos, while red represent mass-Peak Patch. The Eulerian radii were increased by a factor of 3 for visualization purposes. We see that halos generally have the same positions and size (mass) in both methods across all redshifts, and that the large scale structure is very well-reproduced. In Lagrangian space we also show the positions of all N-body particles which end up in halos as the diffuse black background. We see that even N-body halos with an aspherical Lagrangian shape are generally well matched by the mass-Peak Patch halos. Shown are all N-body halos with a mass greater than 100 cells of the simulation, $\sim$3$\times$10$^{13}$M$_\odot$, and the equivalent number of mass-Peak Patch halos, with no mass corrections.

The relatively thin slicing of 20 Mpc in the $z$-direction causes some matching halo pairs to have one member in and one member out of the slice, which artificially increases the number of unmatched halos in the visualization. As well, halos slightly above the minimum mass boundary can have a partner of slightly lower mass, which also increases the visual discrepancy. Therefore, in the following sections we quantify the accuracy of the mass-Peak Patch method in Lagrangian and Eulerian space with a number of halo summary statistics: halo matching, halo clustering, and the halo mass function.

\subsection{Halo Matching}
\label{sec:lag_partner}
\begin{figure}
\begin{center}
\includegraphics[width=1\columnwidth,trim={0 0 0 0},clip]{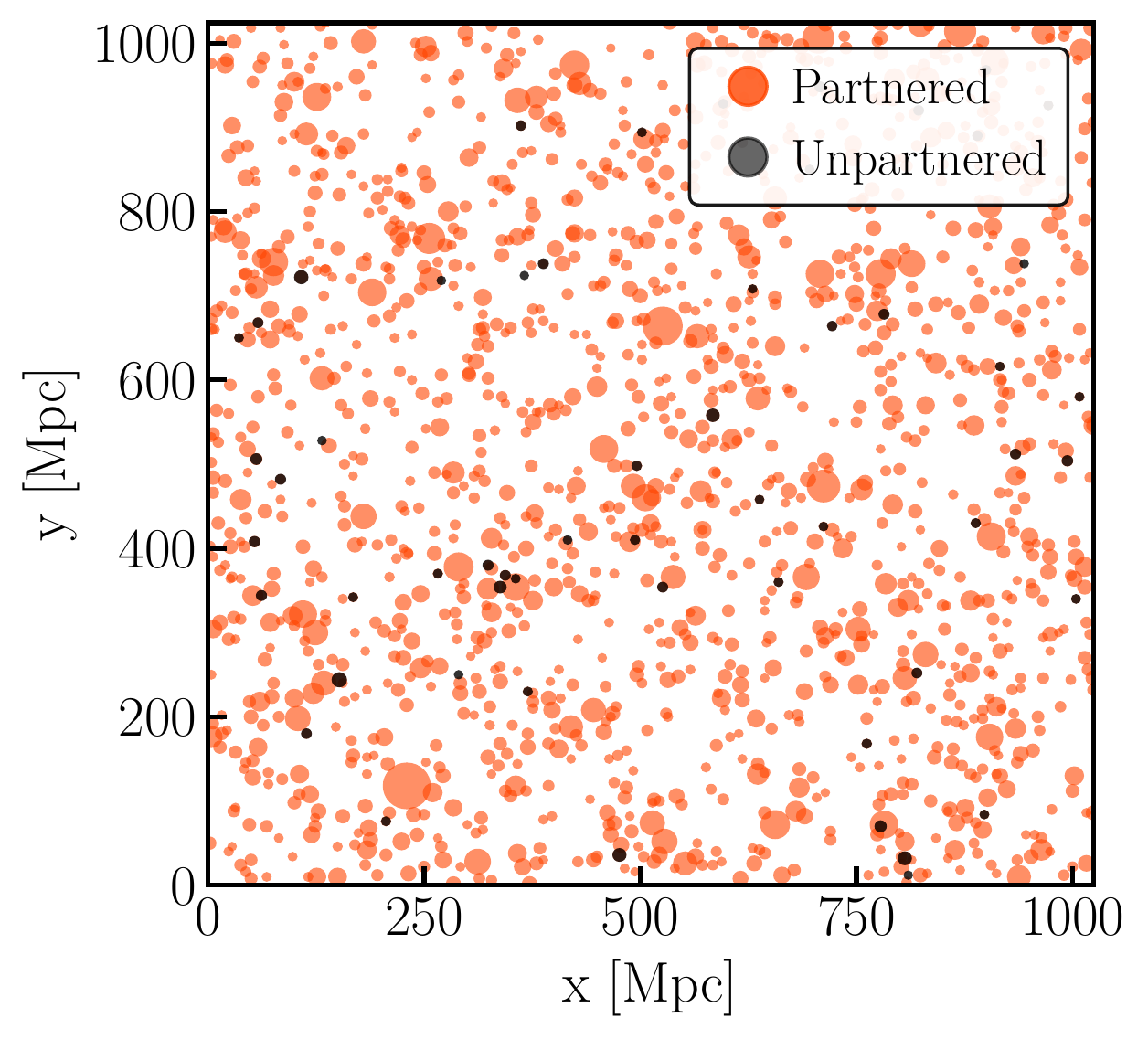}
\includegraphics[width=1.\columnwidth,trim={0 0 0 0},clip]{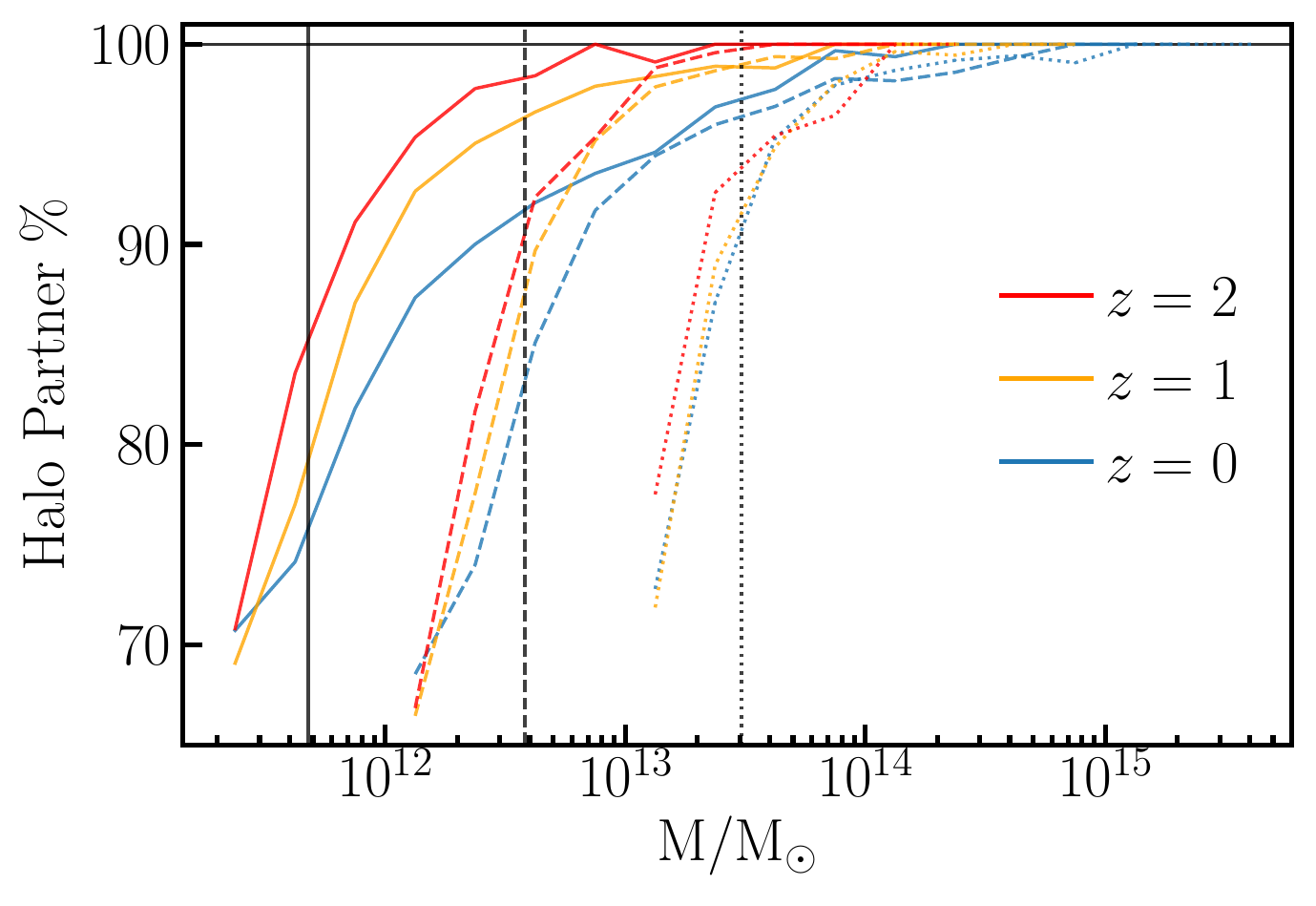}
\end{center}
\vspace{-0.3cm}
\caption{{\bf{Top:}} Partnered (red) and un-partnered (gray) Lagrangian mass-Peak Patch halos in a 20 Mpc thick simulation slice at z=0. {\bf{Bottom:}} Percentage of mass-Peak Patch halos with a Lagrangian N-body partner as a function of halo mass at $z=0$ (blue), $z=1$ (yellow), and $z=2$ (red). The linestyle indicates the resolution of the simulation, where solid, dashed, and dotted denote a 0.5 Mpc, 1 Mpc, and 2 Mpc resolution, respectively. Shown is the mean of the 3 simulation seeds. The vertical lines denote the mass of 100 particles for each simulation resolution.}
\label{fig:halo_partners}
\end{figure}

To quantify the small discrepancies seen between the methods in Figure~\ref{fig:halos}, many of which can simply be a result of the visualization method as described in Section~\ref{sec:halo_visual}, we performed a quantitative analysis of the halo spatial matching between the two methods in Lagrangian space as a function of the mass of a halo. We defined a mass-Peak Patch halo to have an N-body partner if its Lagrangian center of mass is within the Lagrangian radius of an N-body halo; meaning that the two halos are constructed from roughly the same particles of the simulation. This allows for each mass-Peak Patch halo to classified as \textit{partnered} or \textit{unpartnered}. This test is a very efficient post-processing, just involving one more run of the exclusion merger program on the combined catalogues. 

Figure~\ref{fig:halo_partners} shows a visualization of the halos with and without partners (top), and the \textit{halo partner percentage} as a function of the mass of a mass-Peak Patch halo (bottom). The top panel shows the same 20 Mpc thick simulation slice at $z=0$ as Figure~\ref{fig:halos}, where we now instead show mass-Peak Patch halos as two different colours: partnered (red) and unpartnered (gray). We see that a significant fraction of the halos directly correspond to those of the N-body simulation. It is also clear that the only non-partnered halos are those near the low mass end of the simulation, and that every massive halo seen has a partner.

The bottom panel shows the halo partner percentage as a function of mass for the three simulation resolutions and three redshifts. For large halo masses the partner ratio is very close to unity for all simulation resolutions and redshifts, as expected from the top panel. The linestyle indicates the resolution of the simulation while the colour shows the redshift of the halo catalogue. Solid, dashed, and dotted lines denote a 0.5 Mpc, 1 Mpc, and 2 Mpc resolution, while red, yellow, and blue denote redshifts of $z=2$, $z=1$, and $z=0$, respectively. As the halo mass decreases to the mass of 100 particles of the simulation (vertical black lines), which is the smallest halo we expect to accurately resolve, the partner ratio decreases to 90\% for the 2 Mpc resolution simulation for all redshifts, while for the 1 Mpc (0.5 Mpc) resolution simulations we see a variation as a function of redshift from $\sim 90 \%$ at $z=2$ to $\sim 80 \%$ at $z=0$ ($\sim 80 \%$ at $z=2$ to $\sim 70 \%$ at $z=0$) for a 100 particle mass halo. For all simulation resolutions and redshifts we find a high partner percentage of $\geq 90\%$ for halos above a mass of $\sim$200 particles in the simulation.

\subsection{Halo Clustering}
\label{sec:clustering}
The 2-point and 3-point clustering of halos above various mass cuts in simulations are commonly used quantifications of the accuracy of approximate methods when compared to full N-body simulations. Recently, mass-Peak Patch clustering results were presented in a series of three papers \citep{2018arXiv180609477L, 2018arXiv180609497B, 2018arXiv180609499C} which investigated the accuracy of six different approximate methods (HALOGEN \citep{2015MNRAS.450.1856A}, ICE-COLA \citep{2016MNRAS.459.2327I}, log-normal \citep{2017JCAP...10..003A}, mass-Peak Patches, PATCHY \citep{2015MNRAS.450.1856A}, and PINOCCHIO \citep{2013MNRAS.433.2389M}) when compared to N-body for the purpose of creating mock catalogues and covariance matrices for the Euclid collaboration. Also included was the Gaussian recipe of \citep{2016MNRAS.457.1577G} which created covariance matrices only. All methods but the log-normal one simulated 300 independent realizations with matching initial conditions of a 1000$^3$ dark matter particle box, with a side length of 2158 Mpc, to create 300 halo catalogues at $z=1$.

\citet{2018arXiv180609477L} investigated the correlation function results, \citet{2018arXiv180609497B} investigated the power spectrum results, and \citet{2018arXiv180609499C} performed a comparison of the halo bispectrums. When considering abundance matched halos above a minimum mass cut of 100 particles (the only mass sample we provided because of our concerns about patches defined on small numbers of lattice sites), mass-Peak Patch fared well across all halo summary statistics, remaining consistently among the top few methods. Along with ICE-COLA it performed best on the mean and the variance of the power spectrum multipoles, and alongside ICE-COLA and PINOCCHIO it performed best on the mean and the variance of the bispectrum. One interesting result of the bispectrum analysis was that the difference between predictive and abriged particle mesh methods compared to stochastic methods became much greater than for the power spectrum or correlation function, with the exception of PATCHY which fared well. This can partially be attributed to the fact that stochastic methods work by tuning parameters until they can best match specific summary statistics of a few of the N-body runs, such as the correlation function. When then asked to create a halo catalogue to encapsulate a clustering statistic that the method wasn't fit to (such as capturing the correct bispectrum when the parameters of method were fit to give an accurate correlation function), they tend to greatly decrease in accuracy.

This comparison project was very broad in scope, but it was limited to a single simulation resolution of $\sim$2 Mpc, and a single redshift of $z=1$. In this work we expand upon this by looking at three different simulation resolutions and three different redshifts. Although the mass-Peak Patch method is a Lagrangian space halo finder, the final Eulerian state accuracy is generally more important for the desired use of creating large-volume mock halo catalogues to model cosmological observables and their covariances. In the mass-Peak Patch method, halos are moved to their Eulerian positions by calculating the average Lagrangian displacement of every particle belonging to a halo, and for this study the displacements are calculated using second order Lagrangian perturbation theory. Going to an even higher order LPT (at a cost of computational efficiency and memory) has been shown to slightly improve halo displacements in the PINOCCHIO method when compared to N-body \citep{2017MNRAS.465.4658M}. Because of the halo-adaptive smoothing inherent in mPP it is unclear that higher order LPT will be of use, but we leave this for further study:  a modest boost in accuracy may not  be worth the trade off in efficiency. 

\subsubsection{Lagrangian Halo Clustering}
\label{sec:lagclustering}

\begin{figure*}
\begin{center}
\includegraphics[width=1.\textwidth,trim={0 0 0 0},clip]{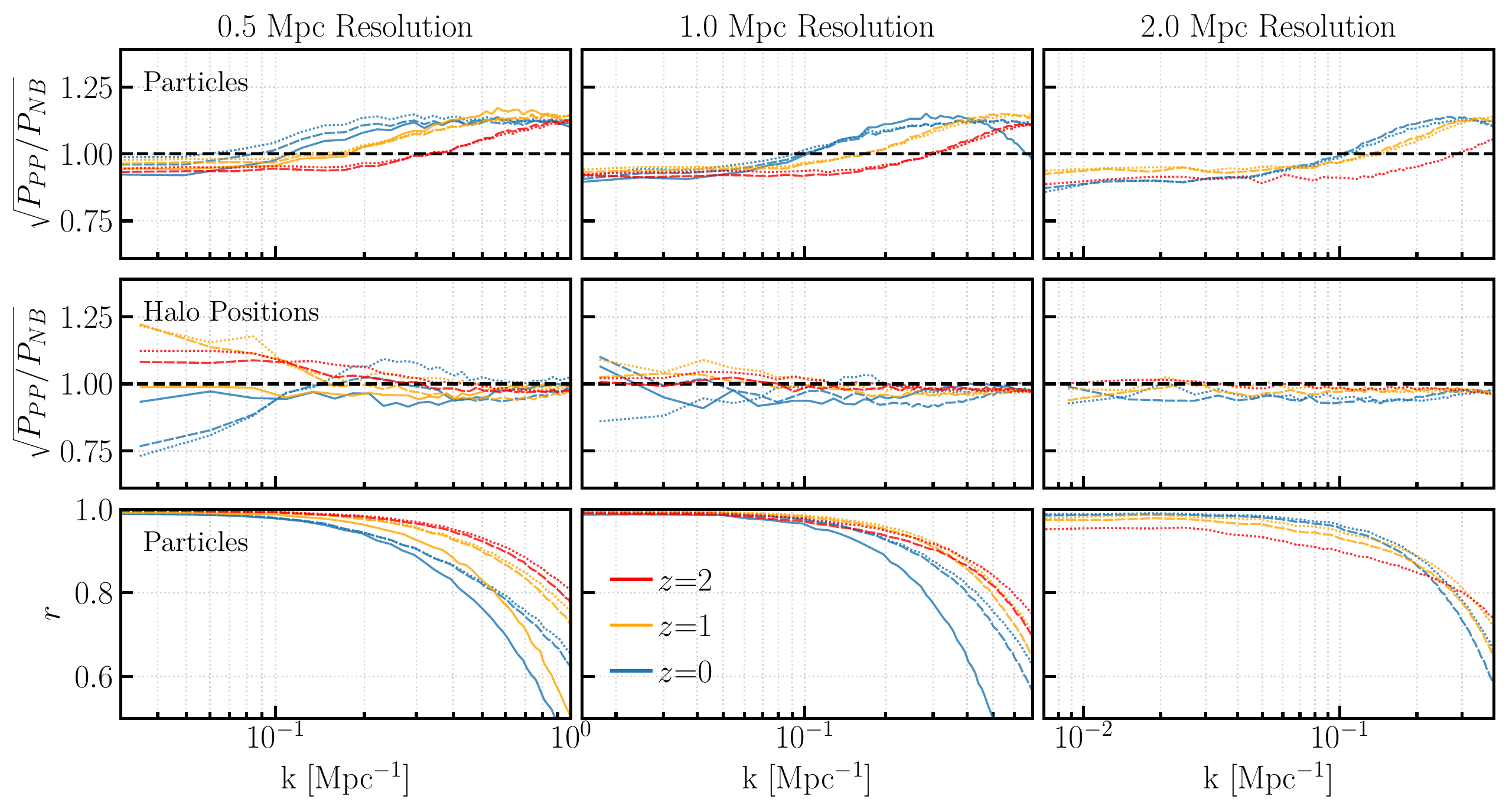}
\end{center}
\vspace{-0.3cm}
\caption{{\bf{Lagrangian Space}}. {\bf{Top row:}} Transfer function of the Lagrangian {\textit{particles that belong to halos}}. {\bf{Middle row:}} Transfer function of the Lagrangian halo positions. {\bf{Bottom row:}} Cross correlation coefficients of Lagrangian particles that belong to halos. The Lagrangian radius of the largest halo expected at $z=[0, 1, 2]$ is roughly R$_h$=[30 Mpc, 15 Mpc, 10 Mpc]. Translating this to a wavenumber results in $k\sim$ [0.1Mpc$^{-1}$, 0.15Mpc$^{-1}$, 0.2Mpc$^{-1}$], which is nearly exactly where we see the deviation of the mass-Peak Patch Lagrangian particle transfer function and particle cross correlation function from the N-body. This is expected, as mass-Peak Patch halos are strictly spherical, while the N-body are not. Additionally, we have not abundance matched the halos in mass, meaning that the total number of particles in mass-Peak Patch halos is not equivalent to N-body in this comparison. Each panel shows the mean results of the three simulation seeds at $z=0$ (blue), $z=1$ (yellow), and $z=2$ (red), for three halo particle number cuts. The linestyle indicates the minimum number of particles per halo, where dotted, dashed, and solid show a 50, 100, and 500 particle minimum, respectively. The conversion from number of particles to halo mass is simply $M_{\rm{min}} = 3.82{\times}10^{12} \left( \frac{\rm{resolution}}{1\rm{Mpc}} \right)^3 \left( \frac{\rm{N_{particle}}}{100} \right) M_\odot$.}
\label{fig:lag_powerspectrum}
\end{figure*}

Figure~\ref{fig:lag_powerspectrum} shows the halo clustering results in Lagrangian space as a function of redshift for the three simulation resolutions. The top row shows the transfer function of the Lagrangian \textit{particles that belong to halos}, the middle row shows the transfer function of the Lagrangian halo positions, and the bottom row shows the cross correlation coefficient of the Lagrangian particles that belong to halos. Each line is the mean of the 3 simulation seeds. The colour indicates the redshift of the analysis, where blue, yellow, and red represent $z=0$, $z=1$, $z=2$, respectively, and the linestyle indicates the minimum number of particles per halo, where dotted, dashed, and solid show a 50, 100, and 500 particle minimum. The conversion from number of particles to halo mass is simply $M_{\rm{min}} = 3.82{\times}10^{12} \left( \frac{\rm{resolution}}{1\rm{Mpc}} \right)^3 \left( \frac{\rm{N_{particle}}}{100} \right) M_\odot$. The lines not shown at $z=2$ are due to the insufficient number of halos found in both methods to perform a meaningful power spectrum calculation.

\begin{figure*}
\begin{center}
\includegraphics[width=1.\textwidth,trim={0 0 0 0},clip]{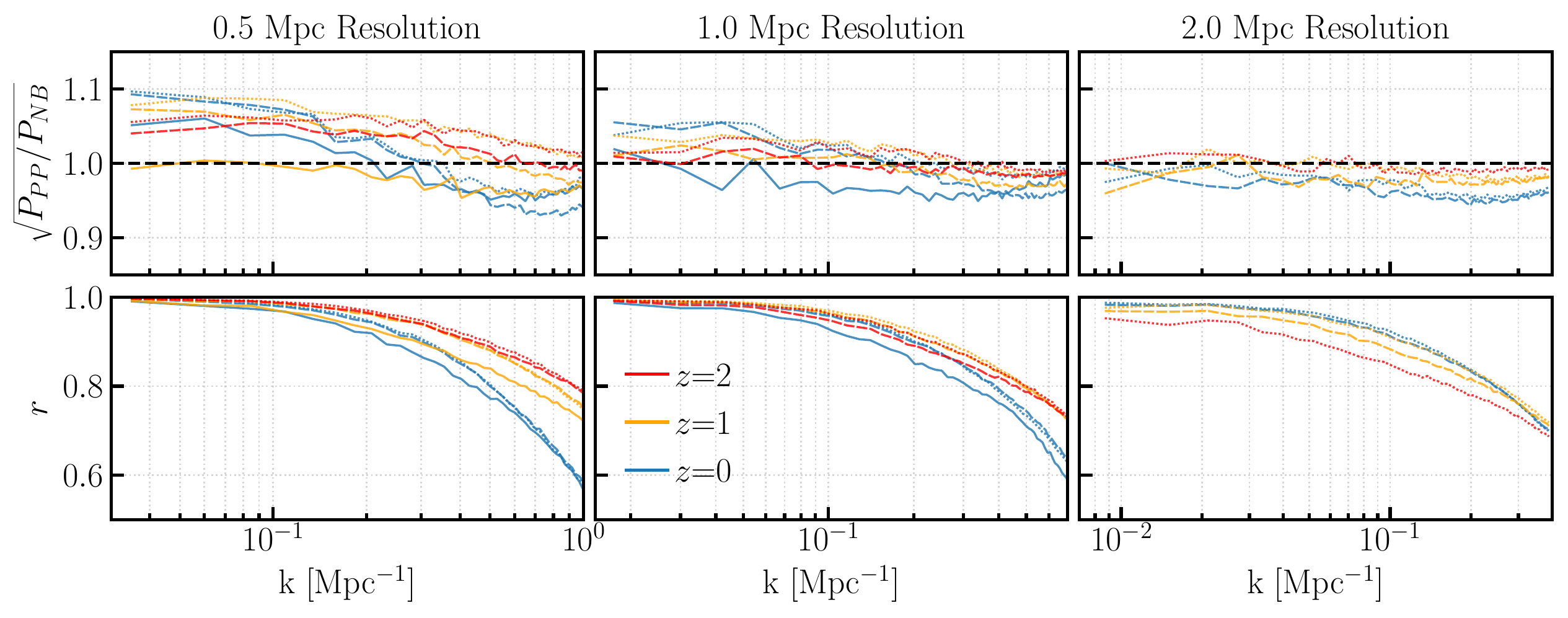}

\end{center}
\vspace{-0.3cm}
\caption{{\bf{Eulerian Space}}. Halo transfer function (top row) and cross correlation coefficient (bottom row) of the N-body and mass-Peak Patch halo catalogues in Eulerian space at $z=[0,1,2]$ for three particle number cuts. Shown is the mean of the three simulation seeds. Each panel shows the mean results of the three simulation seeds at $z=0$ (blue), $z=1$ (yellow), and $z=2$ (red), for three halo particle number cuts. The linestyle indicates the minimum number of particles per halo, where dotted, dashed, and solid indicate a 50, 100, and 500 particle minimum, respectively.}
\label{fig:eul_powerspectrum}
\end{figure*}

For the Lagrangian-space halo particles (top) we find that the transfer function is flat on large scales (small $k$), and within $5-10\%$ for all simulation redshifts, number cuts, and resolutions. On small scales (large $k$) we find that mass-Peak Patch has more power. This is expected since mass-Peak Patch halos are taken to be strictly spherical in Lagrangian space, while N-body halos are strictly spherical in Eulerian space, hence not when mapped back to in Lagrangian space. The Lagrangian radius of the largest halo expected at $z=[0, 1, 2]$ is roughly R$_h$=[30 Mpc, 15 Mpc, 10 Mpc]. Translating this to a wavenumber results in $k\sim$ [0.1Mpc$^{-1}$, 0.15Mpc$^{-1}$, 0.2Mpc$^{-1}$], which is nearly exactly where we see the deviation of the mass-Peak Patch Lagrangian particle transfer function from the N-body. Additionally, we have not abundance-matched the halos in mass, hence the total number of particles in mass-Peak Patch halos is not equal to those in the N-body halos. 

The transfer function for the Lagrangian halo positions (middle) is flat on all scales for the 1 Mpc and 2 Mpc resolution simulations, though with some scale dependence on large scales for the 0.5 Mpc resolution simulation, similar to what is seen in  the Eulerian case. For both the 1 Mpc and 2 Mpc simulation resolutions the transfer function is within $5\%$ for $z=1$ and $z=2$, and within $\sim7\%$ percent for $z=0$. For the 0.5 Mpc resolution simulation we see more deviation at the large scales. When considering $k> 0.1$ Mpc$^{-1}$ we see that for all simulation redshifts, number cuts, and resolutions the transfer remains within $7\%$, but for the 0.5 Mpc resolution simulation it deviates by up to $25\%$ for $k< 0.1$ Mpc$^{-1}$. This may be attributed to the small 256 Mpc size of the box: more investigation is needed.

The cross correlation coefficient of the Lagrangian halo particles (bottom) is close to unity for all simulation redshifts and resolutions at large scales. For the 0.5 Mpc resolution simulation we find, for redshifts $z=[0,1,2]$, that $r\geq 0.95$ at $k$=[0.2Mpc$^{-1}$, 0.3Mpc$^{-1}$, 0.4Mpc$^{-1}$], and that $r\geq 0.9$ at $k$=[0.3Mpc$^{-1}$, 0.45Mpc$^{-1}$, 0.6Mpc$^{-1}$]. For the larger resolution simulations we find a small decrease  at high $k$ values, which is to be expected as these small spatial scales are unresolved.

\subsubsection{Eulerian Halo Clustering}
\label{sec:eulclustering}

Figure~\ref{fig:eul_powerspectrum} shows the halo clustering results in Eulerian space as a function of redshift for the three simulation resolutions. The top row shows the transfer function, while the bottom shows the cross correlation coefficient, where each line is the mean of the 3 simulation seeds. The colour indicates the redshift of the analysis, where blue, yellow, and red represent $z=0$, $z=1$, $z=2$, respectively, and the linestyle indicates the minimum number of particles per halo, where dotted, dashed, and solid show a 50, 100, and 500 particle minimum. The lines not shown at $z=2$ are due to the insufficient number of halos found in both methods to perform a meaningful power spectrum calculation.

We find that the transfer function is nearly flat at large scales for all simulation resolutions, but shows a scale dependence on large scales that increases between z=2 to z=0 for the 0.5 Mpc resolution simulation. For both the 1 Mpc and 2 Mpc simulation resolutions we find that the transfer function is within $5\%$ for all redshifts and number cuts for the entire $k$ range. Considering the 2 Mpc resolution simulation, at $z=[0,1,2]$ the transfer function agreement is within $[5\%, 2.5\%, 1\%]$, and for the 1 Mpc resolution simulation the agreement at $z=[0,1,2]$ is within $[5\%, 2.5\%, 2.5\%]$, across the entire $k$ range. For the 0.5 Mpc resolution simulation the transfer function remains within $10\%$ for all redshifts and number cuts. We also find that the Eulerian-space cross correlation coefficient of the halo positions (bottom) remains high for all resolutions, particle number cuts, and redshifts. The transfer function results of the Euclid comparison project, \citet{2018arXiv180609497B}, a $\sim 2.2$ Mpc resolution simulation at z=1, are nearly identical in amplitude to the results shown in Figure~\ref{fig:eul_powerspectrum} by the dotted yellow line (z=1) of the 2 Mpc resolution simulation, adding further validation to mPP simulations compared with N-body simulations. We find the same exercise for Lagrangian space halo positions reveals more deviation, largely attributable to not adding a mass-dependent position uncertainty: the large halos encompass enormous regions of Lagrangian space. After dynamical evolution, this effect  is partly mitigated, but even for the Eulerian space tests a radius-scaling associated with positional uncertainty  would further improve the halo-halo cross correlation coefficients. One way to do this is to create a 3D map from the catalogue by adding a halo-radius Gaussian uncertainty. The merger test described above uses top hats rather than Gaussians, and uses the halo radius rather than a range of fuzziness scales, but has more statistical content than the power spectra of these 3D fuzzy maps so we have not pursued the fuzzy-halo approach further in this paper. 

\subsection{Mass Function}
\label{sec:hmf}
\begin{figure}
\begin{center}
 \includegraphics[width=1.\columnwidth,trim={0 0 0 0},clip]{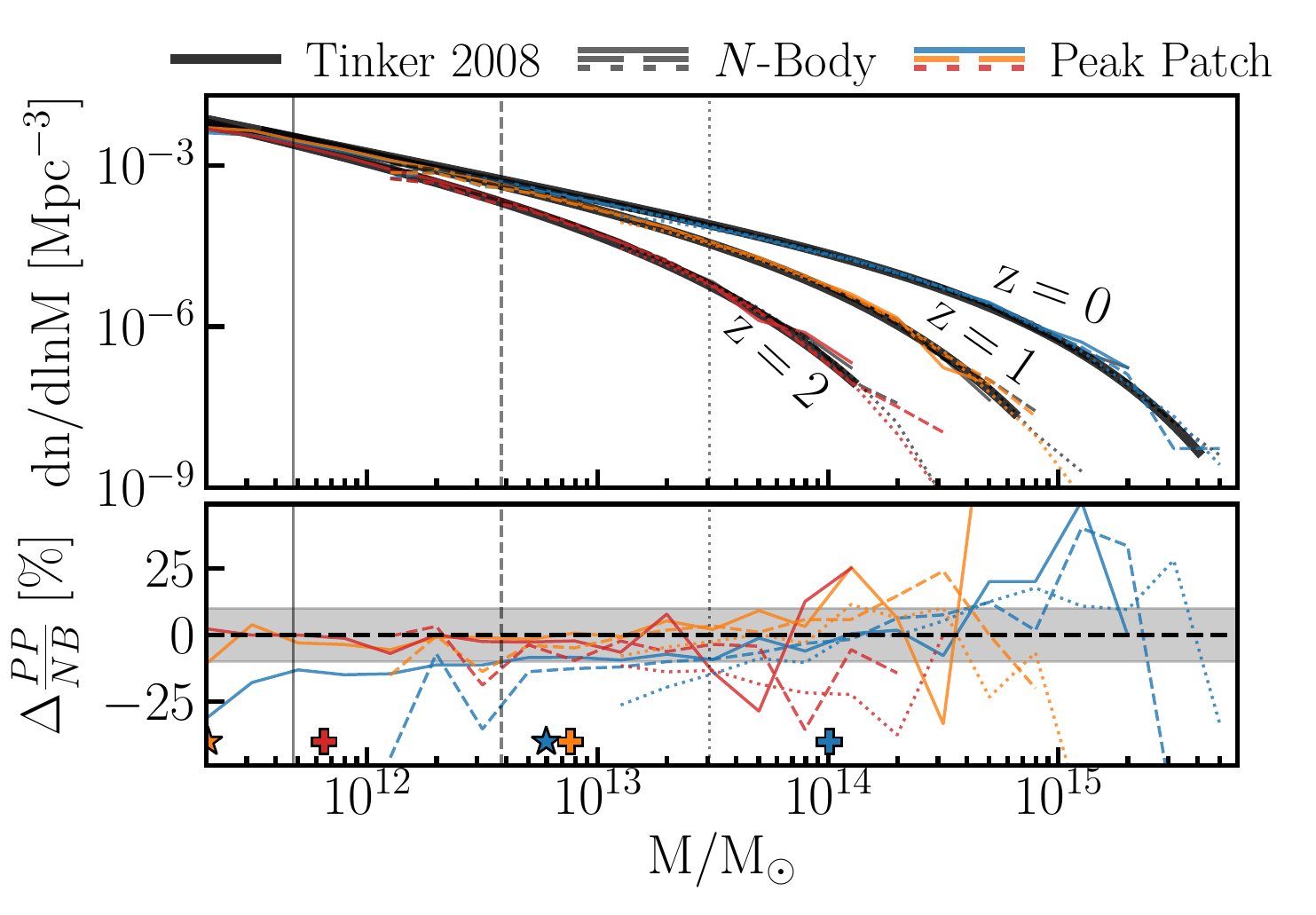}
\end{center}
\vspace{-0.3cm}
\caption{Halo mass functions at $z=[0,1,2]$. Shown is the mean of the 3 simulation seeds for each redshift and simulation resolution. Linestyle indicates the resolution of the simulation, where solid, dashed, and dotted denote a 0.5 Mpc, 1 Mpc, and 2 Mpc simulation resolution, respectively. The thick black lines in the upper panel, which both mass-Peak Patch and N-body accurately match, show the universal mass function of \citet{2008ApJ...688..709T}. The vertical lines illustrate the mass of 100 cells (particles) of a simulation for each simulation resolution, and the gray bar on the bottom panel denotes a $\pm 10$\% deviation of Peak Patch from N-Body. The $\star$ symbols show the value of $M_*$ at each redshift, while the $\boldsymbol{+}$ symbols show the value of $M_{\mathrm{non-lin}}$.  }
\label{fig:box_massfunction}
\end{figure}

Though the halo mass functions are well-reproduced, we note that even if only roughly correct we could still have an accurate method for creating mocks of the large scale structure of the universe. It is straightforward to abundance-match the mass of halos through a few small simulations, or through universal halo mass functions, so the main criterion for accuracy is that the mass-{\textit{ranking}} of the halos of the approximate method matches that of the N-body, and that a sufficient number of halos are detected \citep{2018arXiv180609477L, 2018arXiv180609497B, 2018arXiv180609499C}. The results of the number matched transfer function and cross correlation coefficient of Figure~\ref{fig:eul_powerspectrum} show that this is indeed true for the mass-Peak Patch method.

The halo mass function of the simulations, with no abundance matching or tuning, can be seen in Figure~\ref{fig:box_massfunction}. We show the mean of the 3 simulation seeds, where solid, dashed, and dotted lines denote a 0.5 Mpc, 1 Mpc, and 2 Mpc simulation resolution, respectively, and the colours blue ($z=0$), yellow ($z=1$), and red ($z=2$) indicate the redshift. The vertical lines show where our effective mass cutoff of 100 cells (particles) lie as the resolution varies, and the gray bar on the bottom panel shows the $\pm 10$\% deviation of the mass-Peak Patch from N-Body. We also show the conventional characterization of the mass spectrum $M_*$, defined by $\sigma(M_*)=1.686$,  and the ``non-linear'' mass $M_{\mathrm{non-lin}}$ defined by $\sigma(M_*)=1.686$ ($\sigma(M_{\mathrm{non-lin}})=1.0$, as $\star$ and $\boldsymbol{+}$ symbols, respectively. 

We find that the homogeneous ellipsoidal collapse approximation and binary exclusion of the mass-Peak Patch method produces a halo mass function that agrees with the results from N-body, generally to $10\%$ for all simulation resolutions and redshifts above a minimum halo mass cut of 100 particles. At the highest mass end we see a slightly larger deviation, but this can be attributed to the small number statistics of halos in the largest bins. We also find that mass-Peak Patch reproduces the mass function within $\pm 10\%$, even to below the non-linear scale $M_{\mathrm{non-lin}}$, at all redshifts and simulation resolutions. As in BM, we prefer the Spherical Overdensity approach to Eulerian halo-finding as one more closely matched to what is done in the the mass-Peak Patch algorithm, rather than, e.g., the oft-used friends-of-friends method. More dynamically-refined advanced halo finding based on detailed orbits is not really suitable for comparison with mass-Peak Patch halos.  

\subsection{Additional Validations}
\label{sec:parameters}
Our results on validation are robust across a broad range of simulation resolutions and cosmological parameter choices. Grid cell masses of 2.5 $\times$ 10$^{8} M_\odot$ up to 2.6 $\times$ 10$^{11}M_\odot$ have been used in various simulations, without any significant change to the accuracy of the mass function or power spectrum when compared to a universal halo mass function or an N-body simulation of the same resolution. Flat $\Lambda CDM$ cosmologies with $\Omega_M=(0.2$ to $0.35)$, $n_s=(0.75$ to $0.9)$, and $\sigma_8=(0.7$ to $0.9)$, have also had no qualitative or quantitative effect on the accuracy of the method when compared to N-body. 

We briefly now report on other tests we have done over the years in the lead-up to the current analysis. We compared mass-Peak Patch to spherical overdensity halo catalogues constructed from a suite of simulations created with the CUBEP3M N-body code \citep{2013MNRAS.436..540H}, that covered the cluster formation and galaxy formation range of redshifts shown here, but, more importantly, were extended to cover the early galaxy regime at higher redshifts, where the power spectrum has almost equal power per decade, presenting challenges because of the large cross-talk among quite disparate wavenumbers. A set of mass-Peak Patch and CUBEP3M simulations with box-sizes of 857, 215, and 6.43 Mpc (at $z=10.6$) were constructed from matching initial conditions. These showed halo biasing factors were accurately reproduced. At that time adaptive 1LPT was used to move the halos rather than our now standard 2LPT. Even so, the 6.43 Mpc simulations showed the mass-Peak Patch method was able to reproduce the remarkably deformed spatial structure of large voids at $z=10.6$, in spite of the challenges of nearly-flat density power and extremely large-scale tides. Although in the work presented here we have focused on using simulations at lower redshifts to further our validation, we are confident the mass-Peak Patch method can accurately simulate halo catalogues at high redshifts: the ``first star'' regime and the epoch of reionization.

\section{Discussion \& Conclusions}
\label{sec:discussion}

In this work we presented an updated and massively parallel version of the mass-Peak Patch method of \citet{1996ApJS..103....1B}, an accelerated method for generating simulations of the large scale structure of the universe.  Motivated by the requirements of future large scale structure and CMB surveys and their need for ensembles of large-volume mocks, we have shown that this method (described in Section~\ref{sec:method}), designed to run on massively parallel computing architectures, produces accurate dark matter halo catalogues over 3 orders of magnitude faster than N-body, at a fraction of the memory cost. In Appendix~\ref{sec:appendixa} we discuss the performance and parallelization of the method in detail. 

The mass-Peak Patch method explicitly models nonlinear halo collapse and exclusion. As such, it does not require calibration with respect to N-body simulations, since the basic ingredients of coarse-grained local ellipsoidal dynamics into the nonlinear regime, hierarchical exclusion of small-scale entities within larger-scale entities, and bulk motion and flow, are invariant. Radically varying cosmological parameters does not alter the algorithm, nor does focusing on different halo definitions. This allows modular improvements to be straightforwardly incorporated, e.g., improving flow dynamics beyond 2LPT, adding non-CDM forces to ellipsoidal dynamics, or extending to all manner of beyond the standard model of cosmology situations. That does not mean that there are not fun challenges, e.g., incorporating scale-dependent dark matter components into the ellipsoid collapse model. 

In Section~\ref{sec:comparisons} we compared the accuracy of the method to that of N-body simulations across a broad range of redshifts, $z=[0,1,2]$, and simulation resolutions, [0.5 Mpc, 1 Mpc, 2 Mpc]. We presented a variety of halo summary statistics and showed that, without any parameter tuning, the algorithm produces accurate results. The higher order spatial matching of the mass-Peak Patch halo catalogues compared was investigated in Section~\ref{sec:lag_partner}, the Eulerian and Lagrangian clustering was presented in Section~\ref{sec:clustering}, and the halo mass function results were shown in Section~\ref{sec:hmf}. This follows on from the vintage Bond and Myers validations of the nineties, and sporadic, but instructive, validation exercises subsequently. The most important of those not presented here, in our regard, is the demonstration that the mass-Peak Patch algorithm can successfully treat the tidally extreme regime from which the first small objects emerge at high redshift when compared to CUBEP3M N-body simulations. 

The accuracy of the halo summary statistics presented in this work make the mass-Peak Patch method well-suited for creating large ensembles of dark matter halo catalogues to mock current and next-generation cosmological surveys, ensembles to aid in complex statistical interpretations of the standard model of cosmology and likelihood function creation, and ensembles with cosmic parameters extended to describe beyond the standard model of cosmology. In a forthcoming paper (Alvarez et al. 2018, in prep.) we describe the set of publicly available full-sky extragalactic simulations that we have developed for current and future CMB surveys.\footnote{mocks.cita.utoronto.ca/}
\section*{Acknowledgments}
We thank J.D. Emberson for providing CUBEP3M halo catalogues, and Philippe Berger for helpful discussions.

Research in Canada is supported by NSERC and CIFAR. These calculations were performed on the GPC and Niagara supercomputers at the SciNet HPC Consortium. SciNet is funded by: the Canada Foundation for Innovation under the auspices of Compute Canada; the Government of Ontario; Ontario Research Fund - Research Excellence; and the University of Toronto.

\bibliographystyle{mnras}
\bibliography{main}

\appendix
\section{Parallelization \& Performance}
\label{sec:appendixa}
\subsection{Performance}
\label{sec:performance}

\begin{figure}
\begin{center}
\includegraphics[width=1.0\columnwidth,trim={0 0 0 0},clip]{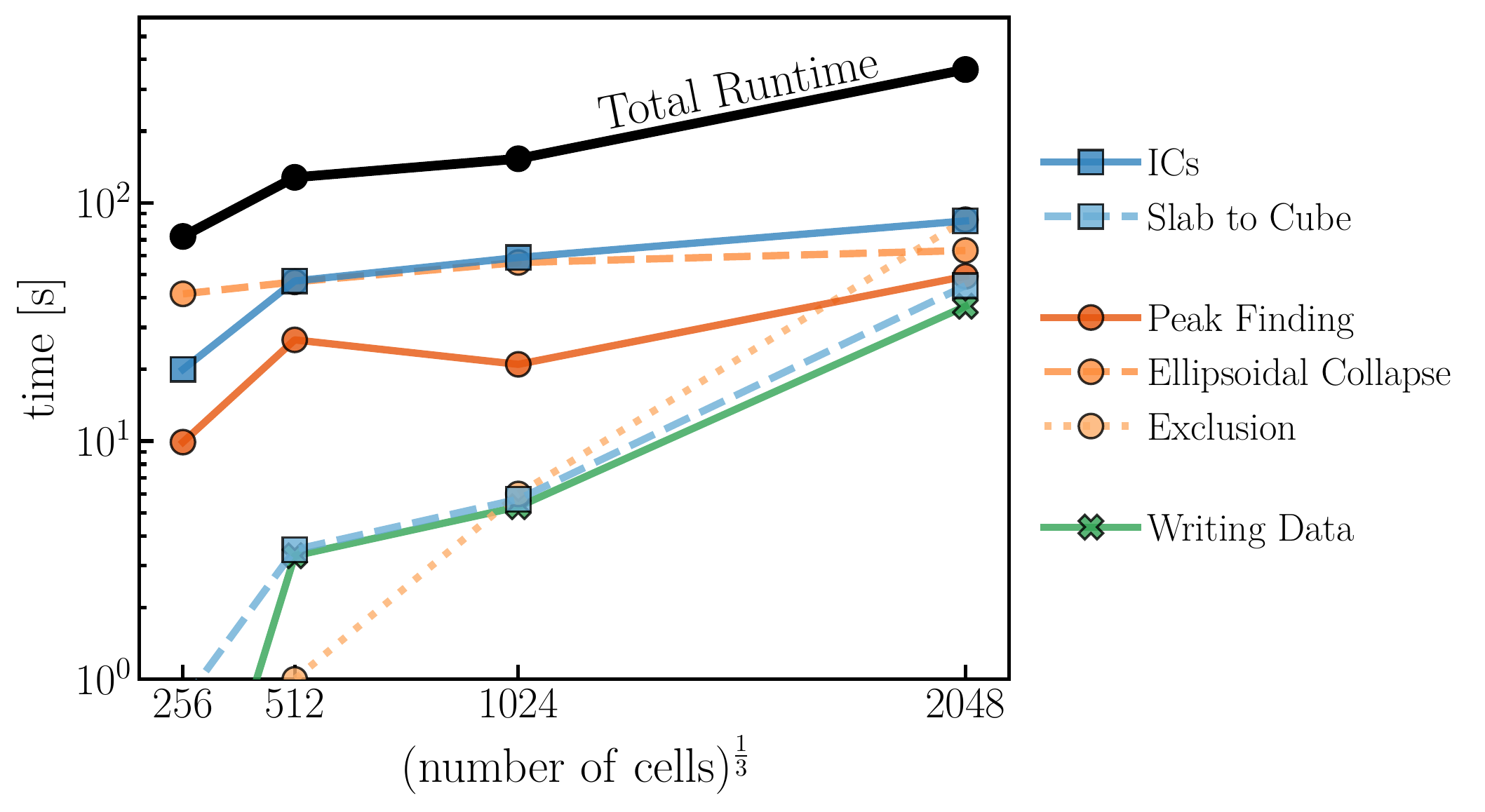}
\end{center}
\vspace{-0.3cm}
\caption{Weak scaling relation of the mass-Peak Patch method on Intel Xeon EE540 2.53 GHz CPU cores of the Scinet-GPC cluster. A fixed lattice resolution of 2 Mpc was chosen, while the number of cells and number of processors was increased accordingly. Each processor computes $\sim$280$^3$ cells representing a volume of $\sim$512$^3$Mpc. The black solid line shows the total time for the simulation. Blue lines are for setting up the initial conditions, orange are for the main halo finding and exclusion algorithims, and green are for the data output. We see that as the simulation size increases, the generation of the initial conditions starts to become the dominating factor, showing the power of the mass-Peak Patch method.}
\label{fig:weakscaling}
\end{figure}

\begin{figure}
\begin{center}
\includegraphics[width=1.0\columnwidth,trim={150 50 150 0},clip]{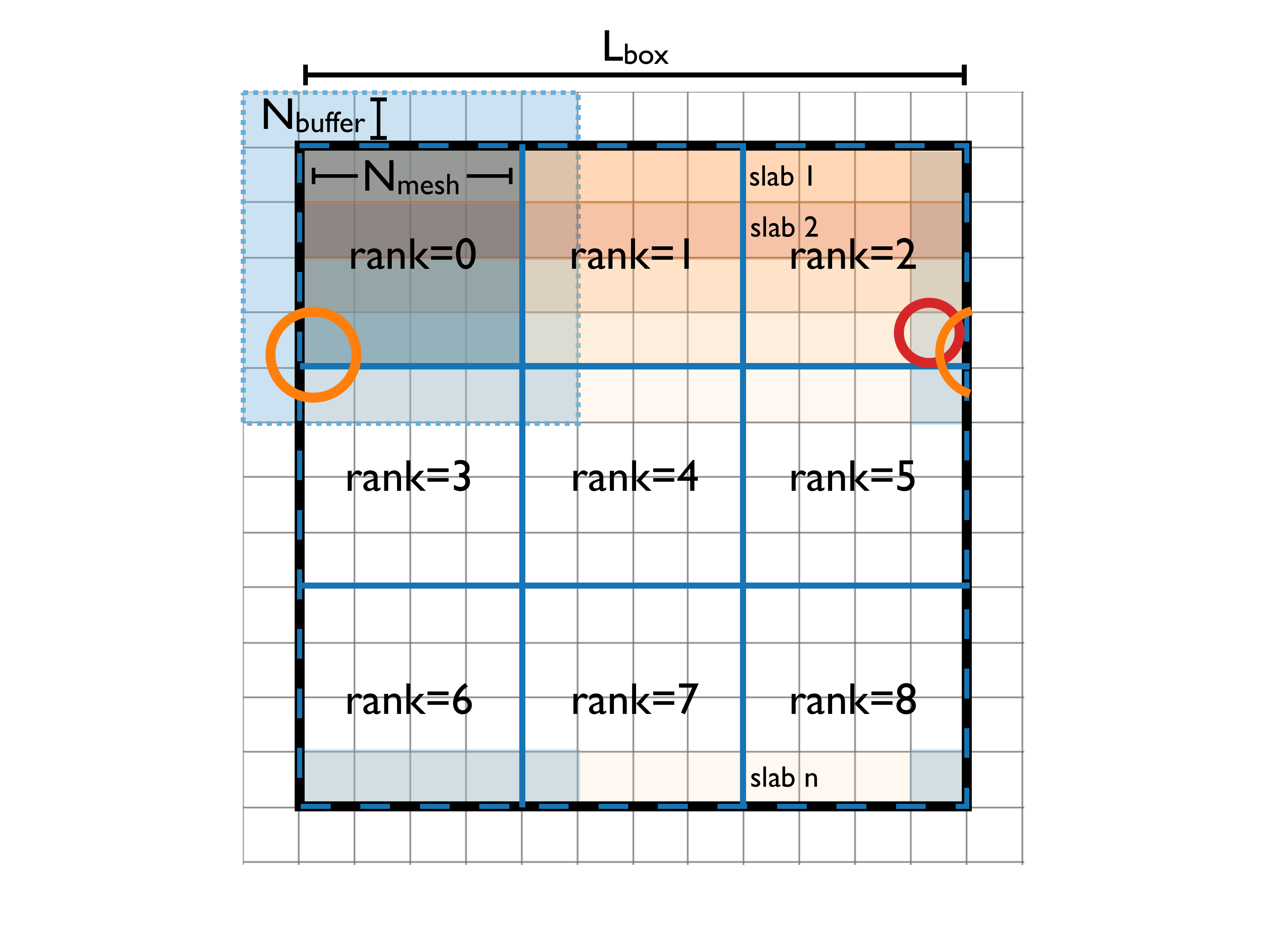}
\end{center}
\vspace{-0.3cm}
\caption{Parallelization setup of the mass-Peak Patch method. The three dimensional initial conditions are created using a slab decomposition, where slabs (shaded in orange) are spread evenly across all MPI processes. Slabs are then re-arranged into cubes, using an additional buffer region (light blue shaded region when considering cube 0) in order to properly perform halo finding on regions near a cube boundary (examples shown by the red and orange halos). MPI processes then independently find halos and calculate their properties, requiring no further message passing until the calculation is complete and halo catalogues are written to disk.}
\label{fig:parallelization}
\end{figure}

\begin{figure}
\begin{center}
\includegraphics[width=0.8\columnwidth,trim={0 0 0 0},clip]{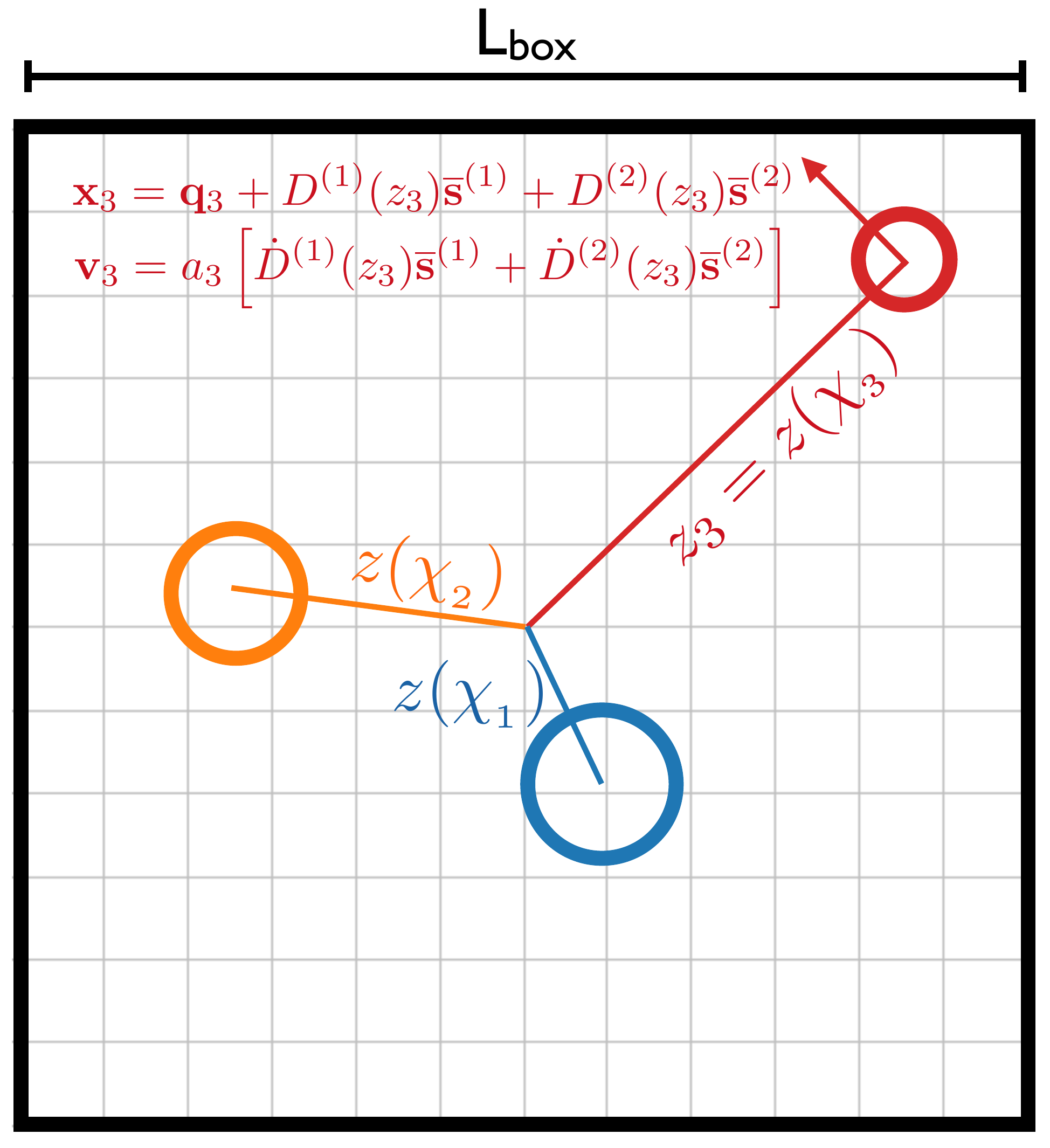}
\end{center}
\vspace{-0.3cm}
\caption{The simplicity of homogeneous collapse allows for light-cone simulations to be created on the fly. The comoving distance from the observer (placed here in the center of the box) to the halo in question ($h_1,h_2,h_3$ in this example) is translated into a redshift, $z=[z_1,z_2,z_3]$. The size of each halo is then determined by calculating the radius of the largest homogeneous ellipsoid that will collapse \textit{by that redshift}, and the first and second order halo displacement vectors are scaled by the appropriate factors of the linear growth factors before moving the halo to its final Eulerian position along the light-cone.}
\label{fig:lightcone}
\end{figure}

Figure~\ref{fig:weakscaling} shows the weak scaling performance of the mass-Peak Patch method. Although we have easily run mass-Peak Patch simulations of a grid size of 6144$^3$ cells, these were unable to be included in the figure due to the setup of the compute cluster used for these tests, which prevented us from keeping the same ratio of processors to simulation volume due to a lack of memory per node. For reference, when comparing to the 4096$^3$ particle Mice Grand Challenge simulation which required $\sim3,100,000$ CPU hours \citep{2015MNRAS.448.2987F,2015MNRAS.447..646C,2015MNRAS.447.1319F}, mass-Peak Patch required a total compute time of $\sim$1,000 hours, a speed up factor of over 3000.
The weak scaling test shows that as the simulation size increases, the generation of the initial conditions starts to become the dominating factor, showing the computational efficiency of the Peak Patch algorithm.                   

\subsection{Parallelization}
\label{sec:Parallelization Scheme}

The initial conditions used for running an N-body or a mass-Peak Patch simulation are in the form of a monolithic cubic periodic grid. As the mass-Peak Patch algorithm needs only semi-local information about neighbouring grid cells to determine the size and location of dark matter halos, this total cubic volume can be divided into subregions with an extra buffer added, each subregion can be passed to a different compute node or (group of) processor(s), and the full calculation can continue without any further message passing. This allows for the simulation to proceed much more rapidly than N-body calculations which require the entire volume to be evolved at the same time, which necessitates complicated message passing.

Figure~\ref{fig:parallelization} shows the domain decomposition of the updated MPI implementation of the mass-Peak Patch algorithm. We create the three dimensional initial conditions using a 1D (or `slab') decomposition, and preform Fast Fourier Transforms using the FFTW3 library\footnote{http://www.fftw.org/}. A slab decomposition has the limitation that, given a cubic simulation of size $N^3$, only $N$ processors can be used to calculate the initial conditions, as each MPI process has to contain at least one slab of data. Although this is a problem for certain types of cosmological simulations, for mass-Peak Patch it is not a limiting step. Each mass-Peak Patch sub-region requires a buffer to allow for halos near the edge of the sub-volume to be properly calculated (shown in Figure~\ref{fig:parallelization} as the light blue shaded region when considering cube 0). This buffer is usually a non-negligible fraction of the total volume, e.g. a $256^3$ region which needs an additional buffer of 32 cells on a side will have a ``usable volume fraction'' of $(256/(256+32{\times}2))^3=0.512$. If we consider the case of $N$ processors for an $N^3$ cubic simulation which we divide into $N$ sub-volumes (in the case of no threading), by increasing the number of processors by a factor of 2$^3$ each sub-volume will decrease to a size of $128^3$, meaning the usable volume fraction drops to $(128/(128+32{\times}2))^3=0.296$, and the buffer starts to dominate the compute resources. Therefore, it is better to utilize shared memory multiprocessing through OpenMP\footnote{https://www.openmp.org/} and use more threads per process in this scenario. For this reason we designed mass-Peak Patch to also use OpenMP for the compute heavy calculations, such as smoothing the density cubes to find density peaks and the homogeneous ellipsoid calculations, and there is no current need to implement a 2D (or `pencil') decomposition. 

The simplicity of homogeneous ellipsoid collapse and the parallelization scheme also allows for light-cone simulations to be created on the fly. This saves a large computational cost when compared to N-body or abridged particle mesh methods, which usually require finely spaced snapshots to be written to disk, halo finding to be performed, and the resulting catalogues to be stitched together. Figure~\ref{fig:lightcone} shows how lightcones are created in the mass-Peak Patch method for no extra computational cost when compared to a single redshift simulation. The comoving distance from the observer to the halo in question is translated into a redshift, and the size of each halo is then determined by calculating the radius of the largest homogeneous ellipsoid that will collapse \textit{by that redshift}. The first and second order halo displacement vector is scaled by the appropriate linear growth factors before moving the halo to its final Eulerian position along the light-cone.

\bsp	
\label{lastpage}
\end{document}